\documentclass[12pt]{article}
\usepackage{graphicx,psfrag,epsf}
\usepackage{enumerate}
\usepackage{url} 
\usepackage{float}
\newcommand{\blind}{0}
\usepackage{graphicx}

\RequirePackage{fix-cm}
\let\accentvec\vec
\makeatletter
\let\@fnsymbol\@arabic
\makeatother

\usepackage[final]{pdfpages}
\usepackage{subfig}
\newcommand{\inprob}{\stackrel{\mathbb{P}}{\rightarrow}}

\usepackage{epsfig}
\usepackage{amssymb}
\usepackage{pifont}
\usepackage[misc]{ifsym}
\usepackage{setspace}
\usepackage{epstopdf}
\usepackage{secdot}
\usepackage{cancel}

\let\vec\accentvec
 \usepackage[T1]{fontenc}
\usepackage{amsmath}
\usepackage{mathtools}
\usepackage{dsfont}
\usepackage{siunitx}
\usepackage{latexsym}
\usepackage{mathrsfs}
\usepackage{color}
\usepackage{authblk}

 \usepackage{amssymb}
 \usepackage{amsthm}
\usepackage{parskip}
\usepackage{sectsty}
\usepackage{times}

\usepackage[noend]{algorithm2e}
\makeatletter
\renewcommand{\@algocf@capt@plain}{above}
\makeatother
\usepackage[colorlinks = true, pdfstartview = FitV, linkcolor = blue, citecolor = blue, urlcolor = blue, bookmarks = false]{hyperref}
\usepackage[authoryear]{natbib}

\newcommand{\m}[1]{\mathrm{#1} }
\renewcommand{\cal}[1]{\mathcal{#1}}
\renewcommand{\v}[1]{\boldsymbol{#1}}
\newcommand{\bb}[1]{\mathbb{#1}}

\newcommand{\cas}{\stackrel{\mathrm{a.s.}}{\rightarrow}}
\newtheorem{remark}{Remark} 

\newtheorem{lemma}{Lemma} 
\newcommand{\dd}{\mathrm{d}}

\newtheorem{proposition}{Proposition}
\usepackage{xcolor, color}
\definecolor{red}{rgb}{0.81, 0.09, 0.13}

\newcommand{\p}[1]{\textcolor{black}{ #1}}

\newcommand{\cmark}{\ding{51}}%
\newcommand{\xmark}{\ding{55}}%
\newcommand{\new}[1]{#1}
\newcommand{\nnew}[1]{#1}

\begin{document}
\def\spacingset#1{\renewcommand{\baselinestretch}%
{#1}\small\normalsize} \spacingset{1}
\if0\blind
{
  \title{\bf Unbiased and Consistent Nested Sampling via Sequential Monte Carlo}

\author[$\dagger$,$\ddagger$,$*$]{Robert Salomone}
\author[$\dagger$,$\ddagger$]{Leah F. South}
\author[$\dagger$,$\ddagger$]{Christopher Drovandi}
\author[$\P$]{Dirk P. Kroese} 
\author[$\mathsection$]{Adam M. Johansen}

\affil[$\dagger$]{School of Mathematical Sciences, Queensland University of Technology, Australia} 
\affil[ ]{}
\affil[$\ddagger$]{Centre for Data Science, Queensland University of Technology, Australia} 
\affil[ ]{}
\affil[$\mathsection$]{Department of Statistics, University of Warwick, England}
\affil[ ]{}
\affil[$\P$]{School of Mathematics and Physics, The University of Queensland, Australia}
\affil[ ]{}
\affil[$*$]{Corresponding author: rsalomone@me.com}

      \date{}
  \maketitle
}
\else {\title{\bf Unbiased and Consistent Nested Sampling via Sequential Monte Carlo} \author{} \date{}
\maketitle
}
\fi

\if0\blind
{
  \begin{center}
\end{center}
} \fi

\vspace{-1cm}
\spacingset{1.1}

\begin{abstract}\normalsize
We introduce a new class of sequential Monte Carlo methods which reformulates the essence of the nested sampling method of Skilling (2006) in terms of sequential Monte Carlo techniques. \nnew{Two new algorithms are proposed, nested sampling via sequential Monte Carlo (NS-SMC) and adaptive nested sampling via sequential Monte Carlo (ANS-SMC)}. \nnew{The} new framework allows convergence results to be obtained in the setting when Markov chain Monte Carlo (MCMC) is used to produce new samples. An additional benefit is that marginal likelihood (normalising constant) estimates \nnew{given by NS-SMC} are unbiased. In contrast to NS, the analysis of \nnew{our proposed algorithms} does not require the (unrealistic) assumption that the simulated samples be independent. We show that a minor adjustment to our ANS-SMC algorithm recovers the original NS algorithm, which provides insights as to why NS seems to produce accurate estimates despite a typical violation of its assumptions. A numerical study is conducted where the performance of \nnew{the proposed algorithms} and temperature–annealed SMC is compared on  challenging problems. Code for the experiments is made available online at 
\if0\blind{\url{https://github.com/LeahPrice/SMC-NS}}\else{\textit{(redacted for blind peer review)}}\fi.
\end{abstract}

\section{Introduction}

A canonical problem in the computational sciences is the estimation of integrals of the form
\begin{equation} \label{eq:canonical}
\pi(\varphi) = \bb E_{\pi}\varphi(\v X) = \int_{E}  \varphi(\v x) \pi(\v x) d \v x,
\end{equation}
where $\pi$ is a probability density on $E \subseteq \bb R^d$, $\mathbf{X}$ is a random variable with probability density $\pi$,  and  $\varphi : E \rightarrow \bb R$ is
a $\pi$-integrable function (note the ``overloading'' of  notation
 for $\pi(\cdot)$, depending on whether the argument is a function $\varphi$ or a
 vector $\v x$).
In Bayesian computation, which is the focus of this work,
$\pi(\v x) = \gamma(\v x)/\cal Z = \cal L(\v x) \eta (\v x) /\cal Z$ where $\pi$ is the {\em posterior} probability
density, $\eta$ is the {\em prior} probability density, $\cal L: E \to \mathbb{R}_{\ge 0}$ is the {\em likelihood}
function, and $\v x \in E$ represents a {\em parameter}. 
Another quantity of interest is the normalising constant 
$
\cal Z = \int_{E} \cal L(\v x)\eta(\v x) \, \dd \v x,	
$
which, in the Bayesian context, is called the {\em marginal likelihood} (or {\em model evidence}) and is often used in model selection \new{(see \cite{fong2020marginal} for benefits and drawbacks of using marginal likelihood for model selection)}. 

Arguably the most popular methodology for estimating \eqref{eq:canonical} is to use {\em Markov chain Monte Carlo} (MCMC). Here, an ergodic Markov chain with $\pi$ as its invariant density is simulated, yielding samples approximately from $\pi$ after a suitably long duration known as the burn-in period. The empirical distribution of these samples can then be used to estimate \eqref{eq:canonical}. For more details, see \citet[Chapters 6--12]{Robert2004}.  Despite the success of MCMC, it can have difficulties in exploring posteriors that have complex landscapes or are multi-modal.  This has motivated the development of population-based methods such as nested sampling and sequential Monte Carlo, where a single chain is replaced by a cloud of samples.

{\em Nested sampling} (NS; \citet{Skilling2006}) is a hybrid Monte
Carlo and numerical quadrature method proposed initially for the
estimation of marginal likelihoods, which also provides estimates of
$\pi(\varphi)$ without requiring additional likelihood evaluations. The method is based on maintaining an ensemble of sample points, and generating new points from 
progressively constrained (nested) versions of the prior.  \new{NS as originally derived has the key property that it reframes a typically high-dimensional integral in terms of a one-dimensional one (see also \cite{birge2012split} and \cite{polson2014vertical} for extensions and generalizations of this approach).}  \new{It} has
achieved wide-spread acceptance as a tool for Bayesian computation in
certain fields, being particularly popular in astronomy (e.g., \cite{Vegetti2009} and \cite{Veitch2015}) and more generally
as a computational method in physics (e.g., \citet{Baldock2017} and
\cite{Murray2005}). \new{For a comprehensive overview of the literature on NS methods,  we refer to the surveys by \cite{ashton2022nested} and \cite{buchner2023nested}, the latter of which includes discussions surrounding the scaling of NS with problem dimension.}

On the other hand, {\em sequential Monte Carlo} (SMC) is a general
methodology that involves traversing a population of particles through
a sequence of distributions, using a combination of mutation,
correction, and resampling steps. SMC has a rich theoretical basis, as
algorithms in this class can be analysed through the theory of interacting particle approximations to a
flow of Feynman-Kac measures. For an introduction to such theory, we refer the interested reader to the comprehensive introductory monograph of \cite{chopin2020introduction}. The use of SMC methodology in a statistical setting began with the {\em Bootstrap Particle Filter} of  \citet{Gordon1993} for online inference in hidden Markov models, and has been the topic of much research (see for example, the survey \cite{Doucet2011}). However, SMC methods in general date much further back to the {\em multilevel splitting} method of \cite{Kahn1951} for the estimation of rare-event probabilities, itself still an active topic of research (e.g., \cite{Botev2012}, \cite{Cerou2012}, and \cite{Cerou2016}).
The special case of SMC where all sampling distributions
live on the same space $E$ is discussed in \cite{DelMoral2006}. In
this setting, one can sample from an arbitrary density $\pi$ by
introducing an artificial sequence of densities bridging from an
easy to sample distribution, say $\eta$, to $\pi$. This approach is
often referred to as SMC in the {\em static} setting. A standard way
to bridge the distributions is through tempering of the likelihood. While static SMC
samplers often make use of MCMC moves, they possess advantages over
the pure MCMC approach in that they are naturally parallelisable, can
cope with complicated posterior landscapes such as those containing
multimodality, and have the added benefit of being able to produce
consistent (and unbiased) estimates of the marginal likelihood as a
byproduct.

A key strength of NS is its suitability for problems where defining a sequence of distributions based on likelihood tempering fails, for example when the model exhibits \new{first-order} phase transitions \citep{Skilling2006}. However, despite the apparent similarities between NS and SMC --- such as sequential sampling and the use of MCMC --- NS lacks convergence results and other theory for practical settings. The aim of the present work is to reconcile the apparent similarity between NS and SMC approaches, and develop new NS approaches based on SMC that have beneficial theoretical properties. To that end, this work provides the following contributions:
\begin{enumerate}

\item Methodologically, we propose two new NS-based SMC algorithms, NS-SMC and adaptive NS-SMC (ANS-SMC), that are derived via different mathematical identities than the original NS and consequentially do \textit{not} assume independent samples are produced at each iteration. 

\item Theoretical results are established for NS-SMC and ANS-SMC that leverage and extend existing SMC results. NS-SMC produces consistent estimators and its marginal likelihood estimator is unbiased (Proposition \ref{prop:FixedNSSMC}). Our main theoretical result appears as Proposition \ref{prop:conv}, which establishes consistency results for estimators arising from ANS-SMC. 

\item A numerical study is conducted involving a difficult example that temperature-based methods fail on, as well as a challenging Bayesian factor analysis model selection problem.  

\end{enumerate}

The layout of the paper is as follows: Sections \ref{sec:NS} and \ref{sec:SMC} provide the requisite background regarding NS and SMC, respectively. Section \ref{sec:NS_SMC} introduces NS-SMC. Section \ref{sec:ANSSMC} introduces the adaptive variant of the algorithm (ANS-SMC), and outlines key differences between ANS-SMC and the original NS algorithm. Section \ref{sec:Numerics} provides a collection of numerical experiments comparing NS, its SMC-based approaches, and traditional temperature-annealed SMC in challenging settings. Section \ref{sec:future} concludes the paper.

\section{Nested sampling}\label{sec:NS}

Nested sampling (NS) \citep{Skilling2006} is based on the identity 
\begin{equation} \begin{split}\label{eq:NestedIdentity}
 \cal Z = \int_{E} \eta(\v x)\cal L(\v x)\, \dd \v x = \eta(\mathcal{L}) &=\int_{0}^{\infty} \bb P(\mathcal{L}({\v X})> l)\,{\rm d}l, 
\end{split}
\end{equation}
where $\cal L$ is a function mapping from some space $E$ to $\bb R_{\geq 0}$,
and $\v X \sim \eta$. Note that $\bb P(\mathcal{L}({\v X})> l)$ is
simply the complementary cumulative distribution function (survival function) of the random variable $\cal
L(\v X)$. We denote this survival function by $\overline{F}_{\mathcal{L}(\v X)}$. A simple inversion argument yields
\begin{equation}\label{eq:NSCOV}
	\int_{0}^{\infty} \overline{F}_{\mathcal{L}(\v X)}(l)\, {\rm d} l = \int_{0}^{1}\overline{F}_{\mathcal{L}(\v X)}^{-1}(p)\, {\rm d} p,
\end{equation}
where $\overline{F}_{\mathcal{L}(\v X)}^{-1}(p)$ is the
$(1-p)$-quantile function of the likelihood under $\eta$, i.e., $\overline{F}^{-1}_{\cal L(\v X)}(p) := \sup\{l : F_{\cal L(\v X)}(l) > p\}$ (see e.g., \cite{evans2007discussion}). This simple
one-dimensional representation suggests that if one had access to the
function $\overline{F}^{-1}_{\cal L(\v X)}$, the integral could then
be approximated by numerical methods. For example, for a discrete set
of values, $0 < \alpha_T < \cdots < \alpha_1 < \alpha_0 = 1$, one could compute the
Riemann sum
\begin{equation} \label{eq:RiemannSum}
\sum_{t=1}^{T}(\alpha_{t-1}-\alpha_{t})\overline{F}_{\mathcal{L}(\v X)}^{-1}(\alpha_{t}),
\end{equation}
as a (deterministic) approximation of $\cal Z$. Unfortunately, the quantile function of interest is typically intractable. NS provides an approximate way of performing quadrature such as \eqref{eq:RiemannSum} via Monte Carlo simulation. The core insight underlying NS is as follows. For $N$ {\em independent} samples $\v X^{(1)}, \ldots, \v X^{(N)}$ distributed {\em exactly} according to a density of the form
\begin{equation}\label{eq:restriction}
\eta(\v x; l) := \frac{\eta(\v x)\bb I\{\cal L(\v x) > l\}}{\eta(\bb I_{{\mathcal{L}>l}})}, \quad \v x\in E, \quad l \in \bb R_{\geq 0},\end{equation}
we have that
\begin{equation}\label{eq:NestedBeta}
\frac{\overline{F}_{\cal L(\v X)}\big(\min_k \cal L(\v X^{(k)})\big)}{\overline F_{\cal L(\v X)}(l)} \sim \mathsf{Beta}(N,1).
\end{equation}
Put simply, consider that one has $N$ independent samples distributed according to the prior subject to the samples lying above a given likelihood constraint, and then introduce a new constraint determined by choosing the minimum likelihood value of the samples. This then defines a new region that encompasses a ${\sf Beta}(N,1)$-distributed multiple of the (unconstrained) prior probability of the previous region. With the latter fact in mind, \cite{Skilling2006} proposes the NS procedure that is formally shown in Algorithm \ref{alg:NS} and proceeds as follows. Initially, a population of $N$ independent samples (henceforth called {\em particles}) are drawn from $\eta$. Then, for each iteration $t = 1,\ldots, T$, the particle with the smallest value of $\cal L$ is identified. This ``worst-performing'' particle at iteration $t$ is denoted by $\breve{\v X}_t$ and its likelihood value by $L_t^N$, with the superscript $N$ signifying that this is a random quantity obtained with $N$ particles. Finally, this particle is moved to a new position that is determined by drawing a sample according to $\eta(\, \cdot \,; L_t^N)$. By construction, this procedure results in a population of samples from $\eta$ that is constrained to lie above higher values of $\cal L$ at each iteration.

After $T$ iterations, we then have $\{L_t^N\}_{t=1}^T$. Each $L_t^N$
corresponds to an unknown $\alpha_t$ such that
$L_t^N = \overline{F}^{-1}_{\cal L(\v X)}(\alpha_t)$. Skilling proposes to
(deterministically) approximate the  $\alpha_t$ values by
assuming that at each iteration the ratio
\eqref{eq:NestedBeta} is equal to its {\em geometric} mean. Such a choice yields the 
approximation that $\alpha_t = \exp(-t/N)$. 
This is the most popular
implementation of NS, and the version considered for the
remainder of the paper. \new{However, it is worth noting that \cite{Skilling2006} proposes another variant which simulates uncertainty by randomly assigning $\alpha_{t} = \alpha_{t-1} B_t$, where $B_t \sim {\rm Beta}(N,1)$, at each iteration}. With the pairs $(L_t^N, \alpha_t)_{t=1}^T$ in hand, the numerical integration is then of the form
\begin{equation} \label{eq:RiemannSum2}
{\cal Z}^{N}= \sum_{t=1}^{T}\underbrace{(\alpha_{t-1}-\alpha_{t})L_{t}^N}_{{\cal Z}^{N}_{t-1}}.
\end{equation}
In practice, the number of iterations $T$ is not set in advance, but
rather the iterative sampling procedure is repeated until some
termination criterion is satisfied. A standard approach \citep{Skilling2006} is to
continue until 
\begin{equation} \alpha_t \max_{1 \leq j \leq N} \cal L\left(\v X^{(j)} \right) < \epsilon
\sum_{j=0}^t {\cal Z}^{N}_j, \label{eq:SkillingStopping}\end{equation} where $\epsilon$ is set sufficiently small to attempt to ensure that the error arising from omission of the final $[0,p_T]$ in the quadrature is negligible. To take into account this interval, a heuristic originally proposed by \citet{Skilling2006},  which we call the {\em filling-in} procedure is to simply add
$\frac{1}{N}\sum_{k=1}^N \cal L(\v X^k)$, \new{scaled by the estimated remaining prior mass}, after termination to the final evidence
estimate, though this is somewhat out of place with the general quadrature construction of the algorithm.

In addition to estimates of the model evidence ${\cal Z}$, estimates of
posterior expectations $\pi(\varphi)$, as in
\eqref{eq:canonical},  can be obtained by 
assigning to each $\breve{\v X}_t$ the weight $\breve{w}_t = {\cal Z}^{N}_t$, and using
 \begin{equation} \label{eq:nested_weighted}
\sum_{t=0}^T \varphi\left(\breve{\v X}_t\right)\breve{w}_t \bigg / \sum_{s=0}^T \breve{w}_s,
\end{equation}
as an estimator.
A formal justification for this is given in \citet[Section 2.2]{Chopin2010}, though in essence it is based on the fact that the numerator and denominator of \eqref{eq:nested_weighted} are (NS) estimators of their corresponding terms in the identity
\begin{equation}
\pi(\varphi) = \eta(\varphi {\cal L})\big/ \eta({\cal L}).
\end{equation}
While the estimator \eqref{eq:nested_weighted} bears a striking resemblance to importance sampling (introduced in Section \ref{sec:SMC}) in its use of a ratio estimator and weighted samples, it is not precisely the same thing. 

Despite the elegance of the NS formulation, a considerable drawback is the requirement of generating perfect \textit{and} independent samples from the constrained distributions at each iteration (i.e., Line 8 of Algorithm \ref{alg:NS}). A typical strategy is to simulate from a Markov kernel with invariant distribution matching the present target, initialised at one of the $N-1$ remaining so-called \textit{live} points. This is a simple practical workaround. However, such an iterative procedure produces neither perfect samples from the desired (conditional) targets (see \citet[Section 4]{l2018generalized} for discussion surrounding this somewhat counterintuitive aspect in a related setting), nor points that are independent of the others. Whilst it could be intuitively expected that the cumulative effect of such imperfections would diminish for larger number of points and/or longer MCMC runs, no theory has yet taken this into account. To accomplish the latter and more, the remainder of the present work explores an alternative, yet closely related approach, based on SMC.

\begin{algorithm}[ht] 
  \setstretch{1.4}
  \SetAlgoSkip{}
  \DontPrintSemicolon
  \SetKwInOut{Input}{input}\SetKwInOut{Output}{output}
  \Input{population size $N$, termination parameter $\epsilon$, boolean decision on whether to perform filling-in}
  \lFor{$k = 1$ \KwTo $N$}{draw $\v X^{(k)} \sim \eta$ }{
    \For{$t \in \bb N$}{
  $m \leftarrow$  ${\rm argmin}_{1\le k\le N} \cal L(\v X^{(k)})$
  \tcp*{identify worst-performing particle}
  $L_{t}^N \leftarrow \cal L(\v X^{(m)})$ \\ 
  $\breve{\v X}_{t-1} \leftarrow \v X^{(m)}$ \tcp*{save sample for inference}
 $\breve{w}_{t-1}\leftarrow \big(\alpha_{t-1} - \alpha_t\big)L_{t}^N$\\
 
$\v X^{(m)} \leftarrow$ a sample from $\eta(\, \cdot \,;L_t^N)$ \tcp*{replace
       worst-performing particle}
 \lIf{{\rm Stopping Condition} \eqref{eq:SkillingStopping} {\rm is satisfied}}{$T^N \leftarrow t$ and {\bf break}}  
 }}
 $\mathcal{Z}^{N,T^N} \leftarrow \sum_{t=0}^{T^N-1} \breve{w}_{t}$\\
  \lIf{\rm{filling-in}}{$\mathcal{Z}^{N,T^N} \leftarrow \mathcal{Z}^{N,T^N} +\alpha_{T^N} N^{-1}\sum_{k=1}^N {\cal L}(\v X^{(k)})$} 
 \KwRet evidence estimator $\mathcal{Z}^{N,T^N}$ and weighted samples $\{\breve{\v X}_t, \breve{w}_t/\sum_{z=0}^{T^N-1} \breve{w}_{z}\}_{t=0}^{T^N}$.
  
  \caption{Nested sampling}\label{alg:NS}
  \end{algorithm}

\label{subsec:issues}

\section{Sequential Monte Carlo}
\label{sec:SMC}

We begin with an overview of importance sampling, which is the
fundamental idea behind SMC. Recall that, in our setting, 
$\pi(\v x) \propto \gamma(\v x)$, where $\gamma$ is a known function. 
For any probability density  $\nu$ such that $\nu(\v x) = 0
\Rightarrow \pi(\v x) = 0$, it holds that 
\begin{equation}\label{eq:ISbasic}
  \begin{split}
\pi(\varphi) & 
= \frac{\nu(\varphi w)}{\nu(w)}, 
\end{split}
  \end{equation}
where $w = \gamma/\nu$ is called the {\em weight} function. The above equation suggests that one can draw $\v X^{(1)}, \ldots, \v X^{(N)} \sim \nu$ and estimate \eqref{eq:ISbasic} via
\[\sum_{k=1}^N \varphi(\v X^{(k)}) \underbrace{w(\v X^{(k)})\bigg/\sum_{i=1}^N w(\v X^{(i)})}_{W^{(k)}},\]
where $\{W^{(k)}\}_{k=1}^N$ are the so-called {\em normalised weights}.

SMC samplers \citep{DelMoral2006} extend the idea of importance sampling to a general method for sampling from a sequence of probability densities $\{\pi_t\}_{t=1}^T$ defined on a common space $E$, as well as estimating their associated normalising constants in a sequential manner. 
This is accomplished by obtaining at each time step $t=0,\ldots,T$ a collection of random samples (called {\em particles}) with associated (normalised) weights $\{\v X_t^{(k)}, W_t^{(k)}\}_{k=1}^N$, for $t=0,\ldots,T$, such that the particle approximations
\begin{equation}\label{eq:measures_converge}
	\pi_t^N(\varphi) := \sum_{k=1}^N W_t^{(k)} \varphi(\v X_t^{(k)}), \quad t=0,\ldots, T,
\end{equation}
converge to $\pi_t(\varphi)$ as $N \rightarrow \infty$. The latter property is referred to as the weighted particles {\em targeting} $\pi_t$. In Bayesian inference, a common sequence of probability densities to use is $\pi_t(\v x) \propto \gamma_t(\v x) = \cal L (\v x)^{\delta_t} \eta (\v x)$ where $0=\delta_0 < \cdots < \delta_t = 1$ so that the first target is the prior and the last is the posterior \citep{Neal2001}. This method is referred to as temperature-annealed SMC (TA-SMC).

SMC samplers use reweighting, resampling and mutation to alter  particles targeting $\pi_{t-1}$ to then target $\pi_t$. A simple SMC sampler that uses MCMC in the mutation step is:
\begin{enumerate}
    \item {\bf Reweight.} Particles $\{\v X_{t-1}^{(k)}, W_{t-1}^{(k)}\}_{k=1}^N$ targeting $\pi_{t-1}$ are reweighted to target $\pi_{t}$. The new weighted particle set is $\{\v X_{t-1}^{(k)}, W_{t}^{(k)}\}_{k=1}^N$, where
$W_t^{(k)} = w_t^{(k)} / \sum_{i=1}^N w_t^{(i)}$, for \[w_t^{(k)} = \gamma_t(\v X_{t-1}^{(k)})/\gamma_{t-1}(\v X_{t-1}^{(k)}),\] and $k=1,\ldots,N$.

    \item {\bf Resample.} The particles are resampled according to their weights, which are then reset to $W_t^k = 1/N$ for $k=1,\ldots,N$. A variety of resampling schemes can be used (see for example \citet{Gerber2019}). The simplest is {\em multinomial} resampling, whereby the resampled population contains $C_k$ copies of $\v X_t^{(k)}$ for each $k=1,\ldots,N$, where $(C_1,\ldots,C_N) \sim {\sf Multinomial}(N, (W_t^{(k)})_{k=1}^N))$. \new{We have considered multinomial resampling in this manuscript for simplicity and to facilitate theoretical analysis; in practice one would anticipate that lower variance resampling schemes would lead to better performance at no cost and we would advocate their use \citep{Gerber2019}.}
	\item {\bf Mutate}. The resampled particles are diversified using a $\pi_{t}$-invariant transition kernel to obtain the final particle approximation to $\pi_t$, $\{\v X_{t}^{(k)}, 1/N\}_{k=1}^N$. 
\end{enumerate}

Provided that $\gamma_0(\cdot)$ is appropriately normalised, estimators of the normalising constants at time $t$ are given by $\prod_{i=1}^{t} N^{-1} \sum_{k=1}^N w_i^{(k)}$,
 which, somewhat remarkably, are unbiased (e.g., \citet[Proposition 16.3]{chopin2020introduction}).
  
\section{Nested sampling via SMC}
\label{sec:NS_SMC}

This section considers a new derivation of NS-type algorithms that is based on importance sampling ideas. In future sections, variants of the basic algorithm are constructed that more closely resemble and demonstrate the relationship to the original NS algorithm. The fundamental idea is to directly reformulate posterior expectations with respect to a sequence of likelihood-constrained prior distributions. We begin by defining a sequence of {\em constrained priors},
\begin{equation}\label{eq:constrained}
\eta_t(\v x) = \frac{\eta(\v x)\bb I\{{\cal L}(\v x) > l_t\}}{\underbrace{\eta(\bb I_{\mathcal{L}> l_t})}_{\textstyle \cal P_t}}, \quad t = 0 , \ldots, T,\end{equation}
that are defined according to the {\em threshold schedule},
$l_0 = -\infty < l_1 < \cdots <  l_{T} < l_{T+1} = \infty$. 
One can sample from this sequence of constrained priors using SMC and we will later show that the output of this SMC sampler can be used to approximate quantities with respect to the posterior. Next, consider splitting the posterior into a collection of distributions with non-overlapping support over different likelihood shells,
\[\pi_t(\v x) = \frac{\eta(\v x) \cal L(\v x) \bb I\{ l_{t} < \cal L(\v x) \le l_{t+1}\}}{\underbrace{\int_{E}\eta(\v x)\cal L(\v x) \bb I\{ l_{t} < \cal L(\v X) \le l_{t+1}\} \dd \v x}_{\textstyle \mathcal{Z}_t}}, \quad t =0,\ldots, T.\]
By construction,  $\bigcup_{t=0}^T \{\v x \in E \, : \,  l_{t} < \cal L(\v x) \le l_{t+1}\} = E$, and thus posterior expectations can be written as
	\begin{align}\label{eq:stratified_identity}
  \bb \pi(\varphi) &= \sum_{t=0}^T \pi \big(\bb I_{l_t < {\cal L} \le l_{t+1}}\big)\pi_t(\varphi) =  \sum_{t=0}^T \frac{\mathcal{Z}_t}{\sum_{s=0}^T {\cal Z}_s}\pi_t(\varphi)  =  \sum_{t=0}^T \frac{\mathcal{Z}_t}{\mathcal{Z}}\pi_t(\varphi).
\end{align}

It follows that 
${\cal Z}_t = \cal P_t \eta_t(\cal L \, \bb I_{{\cal L} \le l_{t+1}})$ and that $\pi_t(\varphi) = \eta_t(\varphi\, \cal L\, \bb I_{{\cal L} \le l_{t+1}})/ \eta_t(\cal L\, \bb I_{{\cal L} \le l_{t+1}})$. Thus, 
\begin{equation}\label{eq:NSSMC_estim}
	\pi(\varphi) = \sum_{t=0}^T\frac{{\cal Z}_t}{\left(\sum_{s=0}^T {\cal Z}_s\right)} \frac{\eta_t(\varphi \cal L \, \bb I_{{\cal L} \le l_{t+1}} \,)}{\eta_t(\cal L \, \bb I_{{\cal L} \le l_{t+1}})}.
\end{equation}

\begin{figure}[h]
  \centering
  \includegraphics[width=6.25cm]{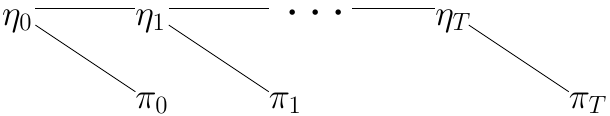}\vspace{2mm}
  \caption{Importance sampling scheme for NS-SMC.} \label{fig:twobranch}
\end{figure}

As a consequence of \eqref{eq:NSSMC_estim}, one need only consider an SMC sampler that sequentially targets $\eta_1,\ldots, \eta_T$, because all terms in \eqref{eq:stratified_identity} can be rewritten in terms of expectations with respect to those distributions. We shall perform SMC sampling along the path of $\eta_t$, while at each step, using the available samples at each iteration to approximate each corresponding $\pi_t$ directly via importance sampling (this branched importance sampling procedure is visualised in Figure \ref{fig:twobranch}). The weighting of those strata are estimated via normalising constant estimates for the constrained posteriors.

Replacing terms in \eqref{eq:NSSMC_estim} with their corresponding SMC estimators, one obtains
\begin{equation}
	\pi^N(\varphi) := \sum_{t=0}^T\frac{{\cal Z}^{N}_t}{\sum_{s=0}^T {\cal Z}^{N}_s} \frac{\eta_t^N(\varphi \cal L \, \bb I_{{\cal L} \le l_{t+1}})}{\eta_t^N(\cal L \, \bb I_{ {\cal L} \le l_{t+1}})}, 
\end{equation}
where each $\mathcal{Z}_t^N =  \cal P_t^N \eta_t^N(\cal L \, \bb I_{{\cal L} \le l_{t+1}})$ and $\cal P_t^N = \prod_{i=0}^{t-1} \eta_i^N(\bb I_{\mathcal{L} > l_{i+1}})$. Noting that
\begin{equation}\begin{split}\label{eq:post_weight}
{\cal Z}^{N}_t \, \frac{\eta_t^N(\varphi \cal L \, \bb I_{{\cal L} \le l_{t+1}})}{\eta_t^N(\cal L \, \bb I_{{\cal L} \le l_{t+1}})} &= \underbrace{{\cal P}^{N}_t \,\cancel{\eta_t^N(\cal L \, \bb I_{{\cal L} \le l_{t+1}})}}_{\textstyle{\cal Z}^{N}_t}\frac{\eta_t^N(\varphi\cal L \, \bb I_{{\cal L} \le l_{t+1}} \,)}{\cancel{\eta_t^N(\cal L \, \bb I_{{\cal L} \le l_{t+1}})}} = {\cal P}^{N}_t \,\eta_t^N(\varphi \cal L \, \bb I_{{\cal L} \le l_{t+1}}),
\end{split}\end{equation}
it is clear that in order to target the (full) posterior $\pi$,  the $k$-th particle targeting $\eta_t$ at iteration $t$ should be assigned the weight
\begin{equation}\label{eq:NSSMC_weights}
    \breve{w}_t^{(k)} := {\cal P}^{N}_t \cal L(\v X_t^{(k)})\bb I\{{\cal L}(\v X_t^{(k)}) \le l_{t+1}\}.
\end{equation}
In turn, an estimator of $\pi(\varphi)$ is obtained via
\begin{equation} \label{eq:NSSMCweights}
	\pi^N(\varphi) = \sum_{t=0}^T \sum_{k=1}^N \breve{W}_t^{(k)} \varphi(\v X_t^{(k)}), \quad \breve{W}_t^k = \frac{\breve{w}_t^{(k)}}{\sum_{s=0}^T \sum_{i=1}^N \breve{w}_s^{(i)}}.
\end{equation}
Pseudocode for this approach is given in Algorithm \ref{alg:NSSMCfixed}, and
\new{it is shown that, under mild regularity conditions which ensure that the probability that all samples simultaneously fall outside the appropriate level sets of the likelihood can be controlled, it provides an unbiased estimate of the marginal likelihood and consistent (in the sample size) estimates of integrals of bounded functions in the following proposition.}

\begin{proposition}[Unbiasedness and Consistency of NS-SMC]\label{prop:FixedNSSMC}
  Under Algorithm \ref{alg:NSSMCfixed} \new{in the setting where multinomial resampling is used}, 
$\bb E{\cal Z^N} = {\cal Z}$ for any $N \in \bb N$. Moreover, \new{provided that $l_1,\ldots,l_{T}$ is chosen so that $\int \eta(d\v{x}_0)\mathbb{I}\{\mathcal{L}(\v{x}_0) > l_{1})\} > \iota$ and for $t=1,\ldots,T$: $\int \mathbb{I}\{\mathcal{L}(\v{x}_t) > l_{t+1})\} \kappa_t(\v{x}_{t-1},d\v{x}_t) > \iota$ for $\eta_{t}$-almost every $\v{x}_{t-1}$ and some constant $\iota > 0$,}
 as $N\to \infty$, it holds that ${\cal Z}^N \cas {\cal Z}$, and for any bounded, measurable function $\varphi$ that $\pi^N(\varphi) \cas \pi(\varphi)$.  
\begin{proof}
By \citet[Proposition 1 and 2]{Cerou2012}, the sequence $\{\eta_t\}$ with MCMC move steps satisfying the appropriate invariant distribution at each iteration has a Feynman--Kac representation. Applying \citet[Theorem 7.4.2]{DelMoral2004} establishes the unbiasedness of each ${\cal Z}_t^N$, which yields the first result via linearity of the expectation operator and that ${\cal Z}^{N} = \sum_{t=0}^{T^N} {\cal Z}_t^N$. The second and third result follows from applying \citet[Corollary 7.4.2]{DelMoral2004}, \new{noting that the hypothesis of this proposition ensures that condition $\mathcal{B}$ of that result holds}, to obtain convergence of the individual terms appearing in $\pi^N$, and combining this with a continuous mapping argument. 
\end{proof}
\end{proposition}

\begin{algorithm}[ht]
  \setstretch{1.4}
  \caption{NS-SMC}
    \label{alg:NSSMCfixed}
  \SetAlgoSkip{}
  \DontPrintSemicolon
  \SetKwInOut{Input}{input}\SetKwInOut{Output}{output}
  \Input{population size $N$, threshold schedule $l_0 = -\infty < l_1 < \cdots <  l_{T} < l_{T+1} = \infty$} 
  ${\cal P}^{N}_0 \leftarrow 1$, $\mathcal{Z}^N \gets 0$ \\ 
  \lFor{$k = 1$ \KwTo $N $}{draw $\v X_{0}^{(k)} \sim \eta$} 
  
   \For{$t \new{\ = 1,\ldots, (T+1)}$} {
$w_t^{(k)} \leftarrow  \bb I\{\mathcal{L}(\v X_{t-1}^{(k)}) > l_t\}$, \enspace for $k=1,\ldots, N$ \tcp*{\new{reweight} $\eta_{t-1} \rightarrow \eta_t$}
     
   $\breve{w}_{t-1}^{(k)} \leftarrow \cal{P}_{t-1}^N \mathcal{L}(\v X_{t-1}^{(k)})\bb I\{\mathcal{L}(\v X_{t-1}^{(k)}) \le l_t\}$  \enspace for $k=1,\ldots, N$ \tcp*{\new{reweight} $\eta_{t-1} \rightarrow \pi$}
   $\mathcal{P}_t^N\leftarrow \mathcal{P}_{t-1}^N N^{-1}\sum_{k=1}^N {w}_t^{(k)}$ \tcp*{\new{normalising constant estimation (NCE) for $\eta_t$}}
   ${\cal Z}_{t-1}^N  \leftarrow  N^{-1}\sum_{k=1}^N \breve{w}_{t-1}^{(k)}$ \tcp*{\new{NCE for $\pi_{t-1}$}}
   ${\cal Z}^{N}\leftarrow {\cal Z}^{N} + {\cal Z}_{t-1}^N $ \tcp*{\new{increment NCE for $\pi$}}
   \lIf*{${\cal P}_t^N = 0$}{$T^N \leftarrow t-1$ and {\bf break} \tcp*{\new{stop if no likelihood $>l_t$}}}
   $\{\widetilde{\check{X}}^{(k)}_{t-1}\}_{k=1}^N$ $\leftarrow$ resample $\left\{\check{X}^{(k)}_{t-1}, w_t^{(k)} \right\}_{k=1}^N$ \tcp*{\new{resample}}  
   Draw $\v X^{(k)}_{t} \sim \kappa_t\left(\widetilde{\v X}^{(k)}_{t-1}, d \v x_t\right)$  for $k = 1,\ldots, N$, where $\kappa_t$ is $\eta_t$-invariant  \tcp*{\new{mutate}}}
 
  \KwRet samples $\{\{\v X_t^{(k)}, \breve{w}_t^{(k)}/ \mathcal{Z}^N\}_{k=1}^{N}\}_{t=0}^{T^{N}}$ and unbiased evidence estimator $\mathcal{Z}^{N}$
 \end{algorithm}

\section{Adaptive NS-SMC}
\label{sec:ANSSMC}

While the first NS-SMC algorithm provided in the previous section is appealing from a theoretical perspective, it is not altogether a practical solution on its own. Generally, one does not have a good idea of a choice of $\{l_t\}_{t=1}^T$ that will perform well. The solution is to determine the thresholds adaptively online, yielding a {\em random} sequence $\{L_{t}^N\}_{t=0}^{T^N}$. A natural approach \citep{Cerou2012} is to specify an {\em adaptation parameter} $\alpha \in (0,1)$, which in turn specifies that $L_{t}^N$ at each iteration is the $\lfloor N(1-\alpha) \rfloor$-th order statistic of the values $\big(\cal L(\v X_{t-1}^{(k)})\big)_{k=1}^N$. It is crucial to note that this will {\em not} guarantee that precisely $\lceil N\alpha \rceil$ particles will lie above each subsequent threshold. For example, even in the case that for any $l \in \bb R_{\geq 0}$, the set $\{\v x : {\cal L}(\v x) = l \}$  is $\eta$-null, a Metropolis-Hastings kernel $\kappa_t(\v x, {\rm d}\v x)$ will have an atom at $\v x$, and thus duplicate particles can be seen in practice. Following \cite{Cerou2016}, the issue is overcome by introducing an auxiliary vector whose elements are independent and identically uniformly distributed on $(0,1)$, and defining the following order on the couples $(\v X^{(i)}, U^{(i)})_{i=1}^N$
\begin{equation}\label{eq:ordering}
\begin{split}
\left(\v X^{(i)}, U^{(i)}\right) <  \left(\v X^{(j)} , U^{(j)}\right) \Longleftrightarrow & \left({\cal L}(\v X^{(i)}) < {\cal L}(\v X^{(j)})\right) \\ & \qquad \quad \text{ or } \left({\cal L}(\v X^{(i)}) = {\cal L}(\v X^{(j)}) \text{ and } U^{(i)} < U^{(j)}\right).    
\end{split}
\end{equation}
In the sequel, order statistics involving the couples is to be interpreted with respect to the above ordering. 
By design, ${\cal P}_t^N = \alpha^t$, and 
thus an online specification of both $\eta_{t}$ and $\pi_{t-1}$ is accomplished where $\alpha$ is the proportion of samples with non-zero weight when reweighting from $\eta_{t-1}$ to $\eta_t$ and $1-\alpha$ is the proportion of samples with non-zero weight when reweighting samples from $\eta_{t-1}$ to the truncated posterior $\pi_{t-1}$. 

Similar to the original NS algorithm, we propose that the termination time is chosen according to an online criteria, and denote it as $T^N$. Specifically, we define 
 \begin{equation} \label{eq:terminationcond}
T^N = {\rm inf} \left\{ t\in \bb N : \frac{{\cal R}_t^N}{{\cal R}_t^N + \sum_{p=0}^{t-1} {\cal Z}_p^N} \le \epsilon\right\},\end{equation}
where 
\[{\cal R}_t^N := \alpha^{t-1}\eta_t^N \left({\cal L}\, \mathbb{I}_{\{\mathcal{L} > L_{t}^N\}}\right), \quad t\in \mathbb{N}. \]
The interpretation of \eqref{eq:terminationcond} is straightforward; the algorithm stops when the estimated remaining proportion of ${\cal Z}$ falls below a specified $\epsilon$. While any $\eta_t$-invariant kernel is suitable for the move step in 
NS-SMC, a stronger condition, matching that of the convergence argument of \citet{Cerou2016}, assumes a particular form for the move step: taking ${\cal A}_L := \{\v x : {\cal L}(\v x) 
\ge L \}$, we insist that for some $L$, the Markov kernels used at 
each iteration $t$ have the form
\begin{equation}\label{eq:kernel_form}
M_{t,L}(\v x, d\v y) = K_{t}(\v x,d\v y) \mathbb{I}_{{\cal A_L}}(\v y) 
+ K_t(\v x,\cal A_L^c)\delta_{\v x}(d\v y),
 \end{equation}
for $x \in \cal A_L$ and $M_{t,L}(\v x, d\v y) = \delta_{\v x}(d\v y)$ 
for $x \not \in \cal A_L$,where $K_t$ is some 
$\eta$-\textit{reversible} Markov kernel on $\bb R^d$.
Practically, any $\eta_t$-invariant Metropolis--Hastings kernel 
(extended to the complement of $\mathcal{A}_L$ as a singular transition at the 
current location) will satisfy this condition (and by the following 
remark, an $r$-fold composition of such kernels will also suffice). With a little additional 
work, the theory could be extended to cover any Markov kernel for which 
the analogue of Proposition 6.1 of Cerou et al. (2016) holds with 
appropriate invariance properties.

\begin{remark}
It is worth noting that $r \in \mathbb{N}$ compositions of the above kernel, denoted $M_{t,L}^{[r]}$, also has the representation in \eqref{eq:kernel_form}. Thus, the forthcoming convergence results hold for arbitrary $r \in \bb N$. 
\end{remark}

Taking into account the above-discussed auxilliary variables, adaptive level thresholds, and termination condition, Algorithm \ref{alg:ANSSMC} is the resulting modified version of Algorithm \ref{alg:NSSMCfixed}. This algorithm is called \textit{adaptive} \nnew{nested sampling via sequential Monte Carlo} (ANS-SMC). 

\begin{algorithm}[ht]
  \setstretch{1.5}
  \caption{Adaptive NS-SMC}
  \label{alg:ANSSMC}
  \SetAlgoSkip{}
  \DontPrintSemicolon
  \SetKwInOut{Input}{input}\SetKwInOut{Output}{output}
  \Input{population size $N$, termination parameter $\epsilon \in (0,1)$, adaptation parameter $\alpha$, number of MCMC repeats $r \in \bb N$}   
  \lFor{$k = 1$ \KwTo $N $}{Draw $\v X_{0}^{(k)} \sim \eta$ and $U^{(k)} \sim {\rm Uniform}(0,1)$} 
  
 \For{$t \in \bb N$}
  { 
  $\v X_{t-1, (k)} \leftarrow$ order statistics of $\left({\cal L}\left(\v X_{t-1}^{(k)}, U^{(k)}\right)\right)_{k=1}^N$ sorted according to \eqref{eq:ordering}  \\ 
  $L_t^N \leftarrow \mathcal{L}\left(\v X_{t-1, (\lfloor N(1 - \alpha)\rfloor}) \right)$
  \tcp*{\new{determine and store threshold}} 
  $\breve{w}_{t-1,(k)} \leftarrow \alpha^{t-1} \mathcal{L}(\v X_{t-1,(k)})$, for $k=1, \ldots, \lfloor N(1-\alpha) \rfloor$ \tcp*{\new{reweight $\eta_{t-1} \rightarrow \pi$}}
  ${\cal Z}_{t-1}^N \leftarrow  N^{-1} \sum_{k=1}^{\lfloor N(1-\alpha)\rfloor} \breve{w}_{t-1,(k)}$ \tcp*{\new{NCE for $\pi_{t-1}$}}
  $ {\cal R_{t}^N} \leftarrow \alpha^{t-1}N^{-1}\sum_{k=\lfloor N(1-\alpha)\rfloor+1}^N  \mathcal{L}(\v X_{t-1, (k)})$ \tcp*{\new{remaining $\pi$ NCE}}
Draw $I_k \sim_{\rm iid} {\rm Uniform}(\{\lfloor N(1-\alpha) \rfloor + 1, \ldots, N\})$ for $k = 1,\ldots, N$ \tcp*{\new{resample}}
  Draw $\v X^{(k)}_{t} \sim M_{t-1, L_t^N}^{[r]}\left(\v X_{t-1, (I_k)}, d \v y\right)$  for $k = 1,\ldots, N$ \tcp*{\new{mutate}}  
  \lIf*{${\cal R}_t^N \left({\cal Z}_0^N + \cdots +{\cal Z_{t-1}^N} + {\cal R_t^N}\right)^{-1} \le \epsilon$}{ $T^N \leftarrow t$ and {\bf break} \tcp*{\new{stopping rule}}} 
 }  \vspace{2mm}
    $\breve{w}_{T^N}^{(k)} \leftarrow \alpha^{T^{N}} \mathcal{L}(\v X_{t}^{(k)})$, for $k=1,\ldots,N$ \tcp*{\new{reweight $\eta_{T^{N}} \rightarrow \pi$}} 
  ${\cal Z}_{T^{N}}^N  \leftarrow  N^{-1} \sum_{k=1}^N \breve{w}_{T^N}^{(k)}$   \tcp*{\new{NCE for $\pi_{T^{N}}$}}
 ${\cal Z}^{N, T^N}\leftarrow \sum_{t=0}^{T^N} {\cal Z}_t^{N}$ \tcp*{\new{NCE for $\pi$}}
 
 \KwRet samples $\{\{\v X_t^{(k)}, \breve{w}_{t}^{(k)}/{\cal Z}^{N, T^N}\}_{k=1}^{N}\}_{t=0}^{T^{N} }$ and marginal likelihood estimator $\mathcal{Z}^{N,T^N}$.\label{alg:NSSMC}
 \end{algorithm}

To formalise the convergence to some fixed stopping time, for $t\in \mathbb{N}$, allow $L_t := F^{-1}_{\mathcal{L}(\v X)}((1-\alpha)^t)$, i.e., the theoretical $(1-\alpha)^t$ quantile of $\mathcal{L}(\v X)$ for $\v X\sim \eta$. Then, let
\[    \xi_t := \eta\left({\cal L}(\v X) > L_{t-1} \right) \int {\cal L}(\v x)\eta(\v x ; L_{t-1}){\rm d}\v x \big/ \mathcal{Z},\] 
and define $T =  \inf\{t \in \mathbb{N}:  \xi_{t} <  \epsilon \}$. The latter is necessarily finite for any $\eta$-integrable ${\cal L}$ (Lemma \ref{lem:terminate} in the supplement). We write $L^2_c(\eta)$ to denote the collection of \textit{continuous} functions that are in $L^2(\eta)$, and $C^1$ to denote the space of continuous functions with continuous first derivatives in all coordinates. With the above notations in hand, the following result establishes \new{that the ANS-SMC algorithm provides consistent estimates of the normalising constant and, also, of the integral of bounded functions (thus characterizing in a weak sense convergence of the sample distribution of the particles to the posterior) to their posterior expectations. The regularity conditions ensure that the likelihood is sufficiently regular to ensure that all quantiles and Monte Carlo expectations behave as intended and that no issues arise in the random stopping procedure.}

\begin{proposition}[Consistency of ANS-SMC]\label{prop:conv}
  \p{Suppose that $\mathcal{L} \in L^2(\eta)$ and is Lipschitz continuous, and that $||\nabla {\cal L}|| >0$, $\eta$-almost everywhere. Then, the quantities associated with Algorithm \ref{alg:ANSSMC} satisfy
  \begin{align}    
    \forall t \in \{1,\ldots,T\}: \mathcal{Z}_{t}^N & \inprob \mathcal{Z}_{t}, \quad  N\rightarrow \infty. \label{eq:Zconv}
  \end{align}
Moreover, provided that $\xi_{T-1} \neq 1 - \epsilon$,  as $N \rightarrow \infty$,
  \begin{align}
    T^N & \cas T < \infty \label{eq:randomtimeconv} \\
    \mathcal{Z}^{N,{T^N}} &\inprob \mathcal{Z} \label{eq:ZTconv}\\ \label{eq:piconv}
    \pi^{N,T^N}(\varphi) &\inprob \pi(\varphi),  \qquad \forall\varphi \in L_c^2(\eta).
  \end{align}}
\end{proposition}
The proof of the above result can be found in the supplement. It involves combining the results of \cite{Cerou2016} with several lemmata addressing the convergence of terms involving indicator variables depending on the random $L_t^N$, as well as taking into account the random stopping time.
\begin{remark}\label{rem:unbiased}
If an unbiased estimator of $\mathcal{Z}$ is desired, \new{the recommended approach (which is also used in our numerical experiments) is to simply} run ANS-SMC (Algorithm \ref{alg:ANSSMC}) first and use the observed $\{L_t^N\}$ as the corresponding sequence $\{l_t\}$ in NS-SMC (Algorithm \ref{alg:NSSMCfixed}).  Thus, an unbiased estimator of $\mathcal{Z}$ can be obtained for approximately a factor of two in the running cost \new{of ANS-SMC. The above is a natural approach, but it is worth noting that other approaches are possible. For example, one could alternatively choose $\{l_t\}$ to be an appropriate subset of the $\{L_t^N\}$ observed from an initial NS run (Algorithm \ref{alg:NS}).} 
\end{remark}
{
\subsection*{Connection between NS and ANS-SMC}  \label{sec:NSvariant}
}
Here, the precise connection between ANS-SMC and NS is established.  
To see that ANS-SMC and NS are actually closely related, consider ANS-SMC with $\alpha = (N-1)/N$. Note that for this particular choice of $\alpha$, we have the identity $
\alpha^t = \alpha^{t-1} - \alpha^{t-1}N^{-1}$. By construction, in the main loop in Algorithm \ref{alg:ANSSMC}, one obtains
\begin{equation} \begin{split}\label{eq:NS_incremental} 
	{\cal Z}^{N}_{t-1} = \alpha^{t-1}N^{-1}\mathcal{L}(\breve{\v X}_{t-1}) = \left(\alpha^{t-1} - \alpha^t\right) {\cal L}\left(\breve{\v X}_t\right),\end{split}
	 \end{equation}
  where $\breve{\v X}_{t-1}$ is the first order statistic with respect to \eqref{eq:ordering}.
The above is \textit{precisely} the same term used at the same point in the NS algorithm, with the exception that \new{we have} $\alpha^t = ((N-1)/N)^t$, \new{in place of} $\alpha_t = \exp(-t/N)$. We refer to the NS algorithm with this alternative choice of weights as ${\rm  NS}^\star$ in the numerical experiments.   \new{Note that $\alpha = (N-1)/N$ is {\em not} the (arithmetic) mean of the compression factor in variable in \eqref{eq:NestedBeta}, which is instead equal to $(N+1)/N$.}

\begin{remark}
As pointed out in \citet[Proposition 2]{Walter2017}, using \new{the values $\alpha^t$ within NS (with perfect and independent sampling) instead of $\alpha_t$} yields the unbiasedness property \nnew{for a NS run with infinite iterations, i.e.,}
$\sum_{k=1}^\infty  \bb E \, {\cal Z}_{t-1}^N  = {\cal Z}$,
whereas using $\alpha_t$ will introduce an overall ${\cal O}(N^{-1})$ bias. \new{The result is arrived at via point-process theory, and is somewhat remarkable in that the estimator still appears to resemble the use of quadrature, yet does not have the introduced bias that one would usually expect from numerical error.}
\end{remark}

At termination, the estimator ${\cal Z}_{T^N}^N$ {\em naturally} recovers precisely the ``filling-in'' heuristic proposed by \cite{Skilling2006} (Algorithm \ref{alg:NS}, Line 10). Further, Algorithm \ref{alg:ANSSMC} (Line 11) reveals that one should use $\alpha^{T^N}{\cal L}(\cdot)$ as the weight function for the final samples. The latter is arguably a minor point from a practical perspective, though is crucially important as the final iteration is necessary for convergence results. We now comment on the most significant difference --- NS replenishes the removed particle at each iteration, while NS-SMC \nnew{and ANS-SMC} employ a resampling and a move step which applies to {\em all} particles. Consequentially, the choice of $\alpha = (N-1)/N$ for ANS-SMC in Algorithm \ref{alg:ANSSMC} is potentially a very wasteful one. Indeed, such a choice makes ANS-SMC precisely $(N-1)$ times more computationally intensive than NS. The following subsection discusses a choice of $\alpha$ that yields an equivalence in terms of computational effort, and provides a discussion as to why moving all particles at each iteration may be beneficial.

\begin{remark}[On the choice $\alpha = e^{-1}$ for ANS-SMC]
The above discussion is instructive in choosing $\alpha$ for ANS-SMC. Note that after $N$ iterations (and hence $N$ sampling procedures), NS will have modified its estimator $p_t$ by a factor of  $ e^{-1} \simeq \left(\frac{N-1}{N}\right)^N$. As ANS-SMC performs $N$ sampling procedures at each single iteration,  to achieve the same effect for the equal computational effort in ANS-SMC, one should choose $\alpha = e^{-1}$.\label{rem:alpha_e1} 
\end{remark}

In light of the above, it is worth noting to derive NS, contrary to the initial derivation in \cite{Skilling2006}, it is {\em not} required by \nnew{algorithms arising from our framework} that the samples be independent at each iteration. Instead, we require only that at each iteration $t$ the empirical measure of the population of particles is a good approximation of the adaptively-chosen target measure $\eta({\rm d}\v x \, ; \, L_t^N)$ at each iteration. 

The effort of applying $M_{t-1}$ to {\em all} particles as in NS-SMC and ANS-SMC should thus be beneficial when the particle approximation is poor and potentially yield improved results for NS-SMC. Conversely, in the setting where the individual MCMC steps mix very well, we may expect NS to exhibit superior performance. This phenomena is explored in the following section. Having unified NS, ANS-SMC, and NS-SMC as members of a larger class of algorithms based around the identity  \eqref{eq:stratified_identity}, Table \ref{table:CompProp} provides a summary of the theoretical properties of the proposed algorithms.

\begin{table}[h]
 \caption{Comparison of algorithm properties between the proposed approaches, and NS under its original formulation.} \label{table:CompProp}
\begin{center}
\footnotesize
\begin{tabular}{c|c| c| c}
 \textbf{Property} & \textbf{NS} & \textbf{ANS-SMC} & \textbf{NS-SMC}  \\ 

 & &  (Algorithm \ref{alg:ANSSMC}) &  (Algorithm \ref{alg:NSSMCfixed})  \\ \hline
Consistent \nnew{(Idealised Case)} & \cmark & \cmark & \cmark \\ 
 &  $N\rightarrow \infty \text{ and } T/N\rightarrow \infty$ &  &  \\ \hline 
Consistent (MCMC) & \fontfamily{cyklop}\selectfont \textit{?} & \cmark & \cmark \\ 
  & & $N \rightarrow \infty$ &  $N \rightarrow \infty, \text{ arbitrary}\  T\in \bb N$ 
  \\ 
  & & (Proposition  \ref{prop:conv}) & (Proposition  \ref{prop:FixedNSSMC}) \\  
  
  \hline   
  Consistent (Random $T^N$) & \xmark & \cmark & NA \\ \hline 
   

Unbiased for $\cal Z$ (MCMC) & \xmark & \xmark & \cmark \\
 & & (See Remark \ref{rem:unbiased})&  (Proposition  \ref{prop:FixedNSSMC}) \\ \hline
 Thresholds Determined Online & \cmark & \cmark & \xmark
\end{tabular}
\end{center}
\end{table}

\nnew{As pointed out by an anonymous referee, the canonical weight choice of $\exp(-t/N)$ tracks the typical behaviour of the remaining amount of prior mass during an NS run in the idealised setting well. A brief discussion surrounding this point as well as a numerical comparison on a series of problems requiring different amounts of prior exploration can be found in Appendix \ref{ap:CompareWeight} of the supplement. In summary, our additional experiments suggest that despite this property, $((N-1)/N)^t$ may be preferable if the goal is to obtain an estimator with lower mean squared error for $\mathcal{Z}$, with the advantage increasing when the bulk of the integral lies in increasingly smaller regions of the prior. However, the experiments demonstrate that $\exp(-t/N)$ may well be a more pragmatic choice in some cases if the goal is to obtain an estimator of $\log \mathcal{Z}$. 
    For this reason, we do not discount the possibility that using NS with weights $\exp(-t/N)$ may potentially be preferred in practice in some settings over NS with weights $((N-1)/N)^t$ or NS-SMC/ANS-SMC. We also note that in our experiment  $\exp(-t/N)$ weights tend to yield an estimator with median value closer to the true value.}

\section{Numerical experiments}
\label{sec:Numerics}

This section presents a series of numerical experiments that explore how the different variants of NS and NS-SMC perform in practice. The first numerical example exhibits a \new{first-order} phase transition. The second example is a highly-challenging factor analysis model selection task.

As the previous sections reason that NS (and hence ${\rm  NS}^\star$), NS-SMC, and ANS-SMC are fundamentally variants on the same type of algorithm, we expect similar performance to a degree. We also compare to adaptive TA-SMC (ATA-SMC), where the temperature schedule is adapted online to maintain a fixed estimate of the effective sample size (ESS) \citep{Jasra2011}, and to standard TA-SMC using temperatures from an independent run of ATA-SMC. \nnew{Convergence results for adaptive temperature-annealed SMC can be found in \cite{Beskos2016}.}

When reporting likelihood evaluation counts, totals for NS-SMC and TA-SMC include the corresponding values from ANS-SMC and ATA-SMC, respectively. \new{We use multinomial resampling in all SMC methods to align with the theory. The results are similar with stratified resampling (Appendix B).}

\subsection{Spike-and-Slab (Phase Transition) Example}\label{subsec:Phase}

A significant advantage of NS-type algorithms is that they possess a particular robustness to certain types of pathologies (see \cite{Skilling2006}). Additional discussion of the robustness enjoyed by NS compared to methods such as Temperature Annealing and Wang-Landau algorithm can be found in the recent article by \cite{partay2021nested}. Such problematic behaviour is often referred to as exhibiting a \new{first-order} {\em phase transition}---models for which the graph of $\log(p)$ vs. $\log \cal L(F^{-1}_{\cal L(\v X)}(p))$ is not concave. In a Bayesian context, a phase transition can be understood intuitively as having a likelihood function that is ``spiked'' and thus increases rapidly in certain regions. While this would seem to be a pathological type of behaviour restricted to problems in computational physics, it is also known to occur in statistical settings, see for example \cite{Brewer2014}.
The following example exhibits such behaviour. 

Write $\phi_{\sigma}$ for the pdf of a multivariate normal distribution with covariance matrix $\sigma^2 \m I$, centred at the origin. Similar to \cite{Skilling2006}, we consider the estimation of ${\cal Z}$ for ${\cal L}(\v x) =  a_1 \phi_{\sigma_1}(\v x) + a_2 \phi_{\sigma_2}(\v x)$ and $\eta(\cdot) $ is the probability density function of the uniform distribution on a unit ball, i.e., $\v x$ such that $||\v x|| \le 1$.  We specify $\v x \in \mathbb{R}^{10}$, $\v \sigma = (0.1, 0.01)^\top$ and $\v a = (0.1, 0.9)^\top$, which introduces a large ``spike'' in $\cal L$ due to the second mixture component. The example considered here is also of particular interest as we are able to perform {\em exact and independent sampling} from each $\eta_t$, (i.e., allows use of the {\em optimal} forward kernel at each iteration within NS-SMC, and samples in a manner satisfying the idealised assumptions within NS).

We also implement a version with MCMC. For an MCMC kernel, we perform ten iterations of a variant of the random walk sampler where we simply propose a movement along a randomly chosen coordinate axis. To ensure the sampler is well suited across progressively narrower densities, at each iteration $h$ is chosen randomly to be either $1/10$ or $1/40$ (with equal probability). We remark that this method strongly outperforms the obvious first choice of the standard random walk sampler. We employ our knowledge of the problem and set the termination criterion to be $L_t^N \ge 0.75 \cal{L}(\v 0)$. Note that while this differs from our standard termination criterion, the convergence is still guaranteed by the convergence of each $L_t^N$ to some deterministic $L_t$ (see also Lemma 4 in the supplement). NS-SMC employs the thresholds obtained via a pilot run of ANS-SMC, and the number of likelihood evaluations from ANS-SMC is also counted in its total. NS and $\text{NS}^*$ use the same number of repeats as NS-SMC, meaning that they use double that of ANS-SMC and double what one would obtain using NS with the same stopping rule. We also implement ATA-SMC for this example, where we use $10$ MCMC repeats, along with the conservative choice of maintaining an ESS of $0.999N$, which will progress slower and allow the particles to move around the space more.
Results for the experiment are given in Table \ref{tbl:sphere10}. 

Two key aspects worth noting are 
(i) For NS, both the variance and bias in the integral estimate seems more pronounced for small $N$ when MCMC is used. The observed (upward) bias for low $N$ is a problem which seems to become more severe when samples are dependent; and (ii)
ATA-SMC {\em fails} on this example. This is unsurprising, as temperature-based methods are ill-suited to such problems. Note also that simple SMC diagnostics can fail to reveal its poor performance (as evidenced by poor standard error estimates). While alternative distribution sequences obtained, e.g., by interpolating to independence \citep{Paulin19}, avoid certain types of phase transitions, there is no general strategy for all such problems.     \new{Finally, we note that \cite[Section 18]{Skilling2006} very briefly mentions a pathological modification of this type of problem that poses a challenge for NS, and the same is true for NS-SMC approaches. In general, given the fundamental similarity between NS and NS-SMC\nnew{/ANS-SMC}, it is inevitable that all NS-based approaches would experience difficulty for similar types of problems. }

\begin{table*}[ht!] \footnotesize
\centering
\caption{ Results for the 10-dimensional spike-and-slab example. Average evidence estimates, standard errors and the average numbers of likelihood evaluations are reported. The analytical ground truth is $\cal{Z} = $ 0.3921 to four decimal places. Estimates have been flagged in red when the null hypothesis of unbiasedness is rejected at a $0.05/30$ significance level so that the familywise error rate is at most $5\%$. The estimate with the lowest mean square error for each combination of sampler and $N$ is emphasised in bold.}
\label{tbl:sphere10}
\begin{tabular}{c|c|c|c|c|c|c|c}
 \multicolumn{2}{c}{}    & \multicolumn{2}{c}{$N=10^2$}  & \multicolumn{2}{c}{$N=10^3$}    & \multicolumn{2}{c}{$N=10^4$} \\ 
 \multicolumn{2}{c}{}    & \multicolumn{2}{c}{($10^4$ repeats)}  & \multicolumn{2}{c}{($10^3$ repeats)}    & \multicolumn{2}{c}{($10^2$ repeats)} \\ 
sampler   & method   & ${\cal Z}^{N}$(SE) & \multicolumn{1}{c|}{evals} & ${\cal Z}^{N}$(SE) & \multicolumn{1}{c|}{evals} & ${\cal Z}^{N}$(SE)       & evals      \\ \hline 
 Exact & NS & \textcolor{red}{0.4532 (0.0026)} & \num{1.0e+4} & 0.3974 (0.0021) & \num{1.0e+5} & \textbf{0.3913 (0.0019)} & \num{1.0e+6} \\ 
 & $\text{NS}^\star$ & \textbf{0.3866 (0.0023)} & \num{1.0e+4} & \textbf{0.3912 (0.0021)} & \num{1.0e+5} & 0.3907 (0.0019) & \num{1.0e+6} \\  
 & ANS-SMC & 0.3953 (0.0033) & \num{5.1e+3} & 0.3942 (0.0028) & \num{5.0e+4} & 0.3931 (0.0027) & \num{5.0e+5} \\ 
 & NS-SMC & 0.3927 (0.0031) & \num{1.0e+4} & 0.3940 (0.0028) & \num{1.0e+5} & 0.3969 (0.0031) & \num{1.0e+6} \\ 
MCMC & NS & \textcolor{red}{0.6235 (0.0234)} & \num{1.0e+5} & \textcolor{red}{0.4136 (0.0067)} & \num{9.9e+5} & 0.4034 (0.0066) & \num{9.8e+6} \\  
 & $\text{NS}^\star$  & \textcolor{red}{0.5346 (0.0203)} & \num{1.0e+5} & 0.4071 (0.0066) & \num{9.9e+5} & 0.4028 (0.0066) & \num{9.8e+6} \\ 
 & ANS-SMC  & \textcolor{red}{0.4720 (0.0081)} & \num{5.0e+4} & 0.4047 (0.0053) & \num{5.0e+5} & 0.3912 (0.0046) & \num{4.9e+6} \\ 
 & NS-SMC  & \textbf{0.3867 (0.0056)} & \num{1.0e+5} & \textbf{0.4030 (0.0050)} & \num{9.9e+5} & \textbf{0.3916 (0.0044)} & \num{9.8e+6} \\ 
 & ATA-SMC & 0.2778 (0.2188) & \num{1.6e+5} & 0.1707 (0.1166) & \num{1.7e+6} & \textcolor{red}{0.0429 (0.0020)} & \num{1.7e+7} \\ 
 & TA-SMC & \textcolor{red}{0.0534 (0.0059)} & \num{3.3e+5} & \textcolor{red}{0.1140 (0.0498)} & \num{3.4e+6} & 1.7353 (1.6824) & \num{3.5e+7} \\ 
\end{tabular}
\end{table*}

\subsection{Factor Analysis}\label{sec:Comparison}
This model choice example considers three target posterior distributions of varying complexity. We consider the monthly exchange rate dataset used in \cite{WestMike1997Bfad}, where exchange rates (relative to the British Pound) of six different currencies were collected from January 1975 to December 1986, for a total of $n=143$ observations. As in \cite{Lopes2004}, we model the covariance of the (standardised) monthly-differenced exchange rates, using a factor analysis model. For $k \le 6$ factors, the data is assumed to be drawn independently from a $\mathcal{N}(\v 0, \Omega)$ distribution, where $\Omega$ can be factorised as $\Omega = \beta \beta^\top+ \Lambda$, for $\beta \in \bb R^{d \times k}$ lower triangular with positive diagonal elements, and $\Lambda$ a diagonal matrix with diagonal given by $\v \lambda \in \bb R_+^d$. The $k$-factor model has $6(k+1)- k(k-1)/2$ parameters, giving 12, 17, and 21 parameters for the one, two and three factor models, respectively. We follow \cite{Lopes2004} and specify the prior distributions as follows:
\begin{equation*}
  \begin{split}
    \beta_{ij} &\sim \cal N(0,1), \quad i < j, \, i=1,\ldots,k, \, j = 1,\ldots, d \\ 
    \beta_{ii} &\sim \cal T\cal N_{(0,\infty)}(0,1), \quad i=1,\ldots,k\\ 
    \lambda_i &\sim {\sf InverseGamma}(1.1, 0.05), \quad i =1,\ldots,d. \\
  \end{split}
\end{equation*}
In order to facilitate improved sampling, we take log-transforms of $\beta_{ii}$ for $i=1,\ldots,k$ and $\lambda_i$ for $i=1,\ldots,6$, which obviates the need to deal with any parameter constraints. The one factor posterior (FA1) is relatively easy to sample from in that the marginal densities are all unimodal. The two factor (FA2) posterior possesses highly separated modes that are challenging to capture for standard MCMC methods (for example, the reversible jump sampler of \cite{Lopes2004} failed to capture this). Finally, the three factor posterior (FA3) contains an exceptionally complex landscape. \new{Plots illustrating the complexity of the posteriors for the two and three factor models are shown in Appendix C of \citet{south2019}.}

\nnew{Our intention is not necessarily to demonstrate the superiority of our proposed method over TA-SMC. Given the variety of possible parameters for SMC (i.e., $N$ and $\alpha$) as well as many possible MCMC kernels, methods of tuning them, and choices for number of MCMC repeats at each iteration, there most likely exists an appropriate choice of these factors for any given problem that will allow one method to outperform the other. Thus, we instead aim to simply make our best efforts using our experience with SMC to get the best out of both algorithms in an automated manner, and observe the results.}

For each of the three target distributions, we execute 100 runs of all algorithms. We use $\alpha = e^{-1}$ for ANS-SMC, as per Remark \ref{rem:alpha_e1}, and we similarly target an ESS of $e^{-1}N$ in ATA-SMC. We use $N=1000$ in NS, ${\rm  NS}^\star$, ANS-SMC and NS-SMC. To maintain a similar total number of likelihood evaluations in ATA-SMC and TA-SMC, we use $N=3000$. We determine the stopping point using $\epsilon = 10^{-5}$ in ANS-SMC and using $\epsilon = 10^{-8}$ in NS and ${\rm  NS}^\star$ to achieve a similar number of likelihood evaluations. At each SMC iteration for the one, two and three factor models, we apply 10, 20 and 30 MCMC steps respectively of a random-walk Metropolis--Hastings sampler\nnew{, arguably the most common proposal choice for both NS and ATA-SMC algorithms}. Following standard practice, we use proposals $\mathcal{N}(\v X, \frac{2.38^2}{d}\widehat{\Sigma})$, where $\v X$ is the current location, $d$ is the dimension of the target distribution, and $\widehat{\Sigma}$ is a covariance matrix. This covariance matrix is estimated from the empirical covariance of the particles for NS, $\text{NS}^\star$,
ATA-SMC and ANS-SMC \citep[see e.g.][]{Jasra2011}. The adapted covariances, levels and stopping points from ATA-SMC and ANS-SMC are used as fixed values in TA-SMC and NS-SMC, respectively.

\begin{figure*}[ht!]
  \centering
   \includegraphics[angle=360, width=0.8\textwidth]{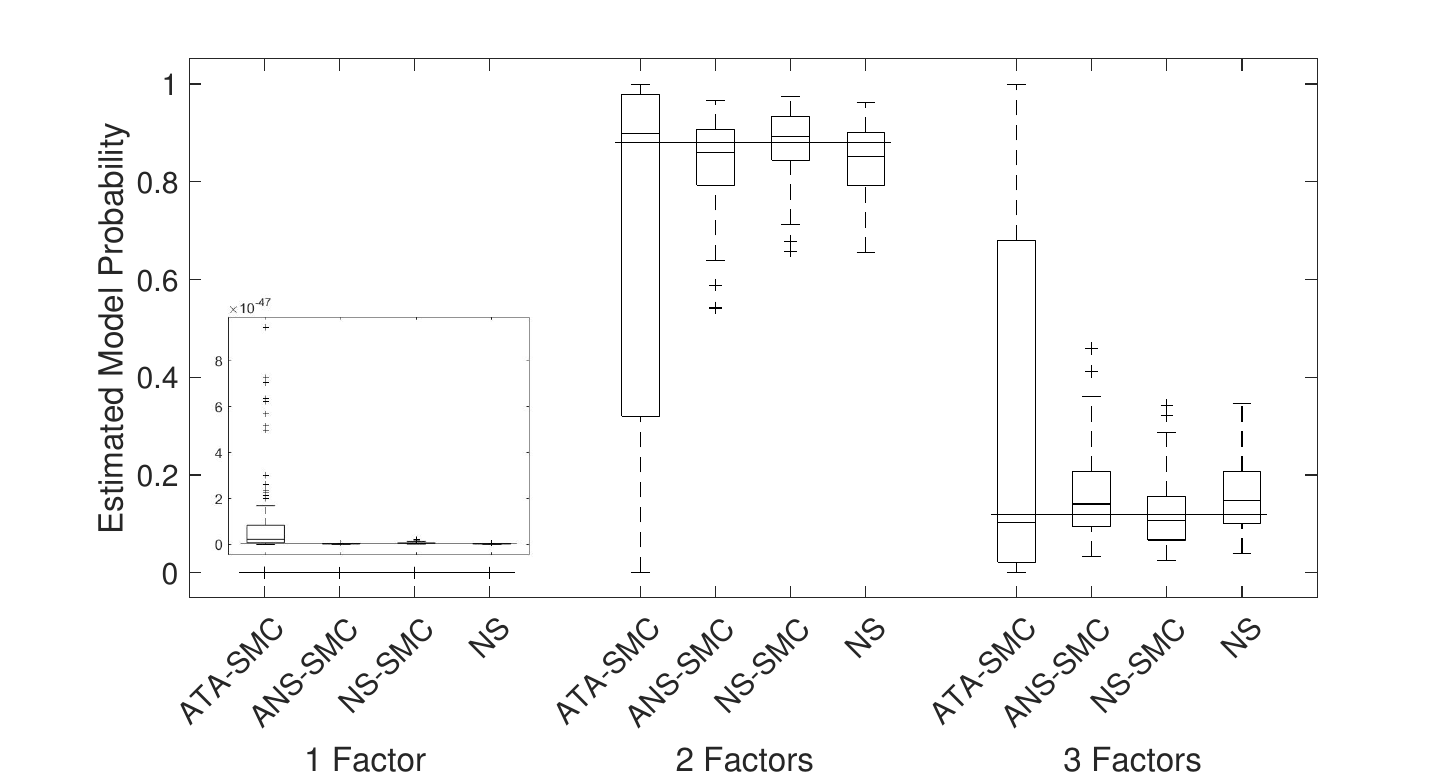}
  \caption{Model probability estimates from 100 runs of all algorithms in the factor analysis example. The straight lines running through the boxes are the estimated model probabilities from the gold standard. }\label{fig:boxFA}
\end{figure*}

\begin{figure}[ht!]
\begin{center}
{\includegraphics[width=0.85\textwidth]{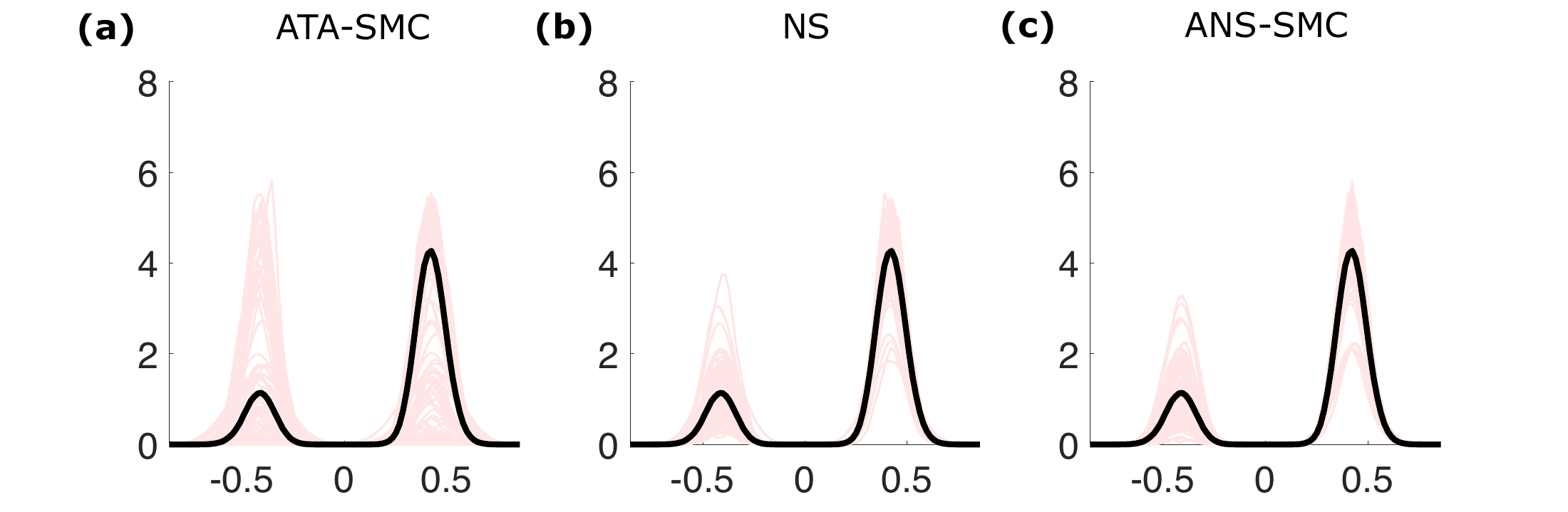}}    
\end{center}
	\caption{Kernel density estimates for the marginal posterior distribution of $\beta_{32}$ in the two-factor model. Estimates from 100 runs of the specified sampler are shown with transparency and the gold standard is shown with a bold line.}
	\label{fig:FA2_marginals}
\end{figure}

Figure \ref{fig:boxFA} displays the estimates for the model probabilities from these 100 runs. As a concise summary of overall sample quality, Table  \ref{tab:FA_KSD} reports the average kernelised Stein discrepancy (KSD) obtained using the inverse-multiquadric (IMQ) kernel with bandwidth parameter set to one. For computational tractability of KSD calculations in NS, $\text{NS}^\star$, ANS-SMC and NS-SMC, $3000$ samples were obtained by resampling the weighted samples. For further details regarding KSD and its use as a measure of sample quality, see \cite{gorham2017measuring}. Figure \ref{fig:FA2_marginals} illustrates kernel density estimates for the runs across different methods. The gold standard shown in Figures \ref{fig:boxFA} and \ref{fig:FA2_marginals} is from an extended ``gold standard" run of TA-SMC with $N = 4\times 10^5$.

Together, the results in the figures demonstrate that, as expected, NS and NS-SMC methods perform similarly, which is to be expected. The NS-based methods also outperform TA-SMC under the experimental settings considered, particularly when the number of likelihood evaluations is taken into account. 

\begin{table}[ht!]
  \centering 
  \caption{Factor analysis average KSD and average number of likelihood evaluations. } \footnotesize
  \label{tab:FA_KSD}
  \begin{tabular}{c|c|c|c}
 Factors & Method & Avg. KSD & Avg. evals\\ \hline 
1 & ATA-SMC  & 1.63  & \num{4.0e5} \\ 
    & ANS-SMC   & 1.35  & \num{3.5e5} \\ 
    & NS-SMC   & 1.45 & \num{6.9e5}\\ 
    & NS    &  1.36   & \num{4.1e5}\\ 
\hline 
2 & ATA-SMC   & 5.66 & \num{8.9e5}\\ 
    & ANS-SMC  & 4.21  & \num{6.8e5}\\ 
    & NS-SMC  & 4.58 & \num{1.4e6}\\ 
    & NS  & 4.14 & \num{7.9e5}\\ 
\hline
3 & ATA-SMC & 6.88   & \num{1.3e6}\\ 
    & ANS-SMC  &  7.37  & \num{1.0e6}\\ 
    & NS-SMC   & 10.19  & \num{2.1e6}\\ 
    & NS         &  7.54  & \num{1.2e6}\\ 
\end{tabular}
\end{table}

\section{Discussion}
\label{sec:future}

An alternate formulation of NS-type algorithms based entirely on Monte Carlo arguments was provided, as opposed to the original formulation which was a hybrid of Monte Carlo and numerical quadrature ideas. The two proposed nested sampling SMC algorithms and the original NS are intimately related under this new derivation. 

\new{The main drawback with nested sampling is its dependence on sampling directly from the constrained prior or treating runs of a Markov chain with that as an invariant \emph{as though they were} such samples.  The SMC construction allows us to construct dependent samples from sequences of distributions in a controlled way so that we can draw conclusions without making use of independence. } NS-SMC and ANS-SMC possess theoretical guarantees in this setting where MCMC is used. In that respect, the two algorithms are the first of their kind.

Given that this work presents a new perspective through which NS-type algorithms can be designed, there are a number of interesting extensions. 
Methodologically, as a pure Monte Carlo approach, NS-SMC \nnew{and related algorithms} might be improved further by variance reduction techniques (e.g., \cite{neufeld2015variance}, \cite{alexopoulos2023variance}, \cite{south2022semi}). The proposed methods could also benefit from advances in the NS and SMC literature. For example, variants that allow for recycling intermediate MCMC steps as in the recently proposed {\em waste free SMC} approach of \cite{wastefreesmc} would likely yield further practical improvements in performance for the same computational effort.  \new{Approaches for {\em posterior repartitioning}, that is, replacing $\eta$ and $\mathcal{L}$ in a manner that leaves their product invariant yet yields improved NS performance \new{\citep{chen2019improving, chen2023bayesian}}, readily extend to NS-SMC approaches.} \new{We also note that one potential application of the unbiasedness of NS-SMC potentially worth exploring is its use within pseudo-marginal MCMC \citep{andrieu2009pseudo}, which requires an unbiased estimator.} \nnew{Particularly, in light of this, it would be interesting to explore whether the approach of \citet{Brehier2016} can be adapted to the NS-SMC context to provide unbiased adaptive algorithms.}.

Finally, it is worth noting that MCMC methods for likelihood-contour constrained spaces remains a relatively unexplored area worthy of new methods, as is methods for effective calibration of MCMC within SMC algorithms. For the latter case, \nnew{methods and theory from \citet{Fearnhead2013} and \citet{Beskos2016} could be extended}. One possible approach involves choosing MCMC kernel parameters that optimise the mean or median estimated expected squared jumping distance \citep{Pasarica2010} via a pilot step at each iteration. Similarly, the number of MCMC repeats within SMC could be determined by observing the sum of the expected jumping distances over a number of pilot steps, and using this to estimate the minimum $r$ that would be required to meet a desired expecting jumping distance. Such heuristics will inevitably introduce bias into the algorithm, although this bias could be removed through performing an NS-SMC run with fixed algorithm hyperparameters.

\if0\blind
{
{\section*{Acknowledgements}}
This work has been supported by the Australian Research Council Centre of Excellence for Mathematical \& Statistical Frontiers (ACEMS), under grant number CE140100049. Leah South is supported by a Discovery Early Career Researcher Award from the Australian Research Council (DE240101190). Christopher Drovandi is supported by the Discovery Program of the Australian Research Council (DP200102101). Adam Johansen acknowledges support from the United Kingdom Engineering and Physical Sciences Research Council (EPSRC; grants EP/R034710/1 and EP/T004134/1) and from United Kingdom Research and Innovation (UKRI) via grant number EP/Y014650/1, as part of the ERC Synergy project OCEAN. Computing resources and services used in this work were provided by HPC and Research Support Group, Queensland University of Technology, Brisbane, Australia. We thank Michael Betancourt, Brendon Brewer, and Zdravko Botev for correspondence and discussions related to NS. We also extend a special thanks  to Christian Robert, whose many blog posts on NS helped influence this work, and played a large part in inspiring it. 
} 
\else 
\fi

\section*{Data Availability}
All datasets used are and were previously publicly available. Source code required to reproduce the experiments is available via \url{https://github.com/LeahPrice/SMC-NS}. 

\section*{Conflicts of Interest}
There are no conflicts of interest.

\bibliographystyle{spbasic}      
\bibliography{SMC} 

\clearpage

\appendix

\section{Proof of Proposition 2 (Consistency of ANS-SMC)} \label{ap:theory} 
Allow $\gamma_p$ and $\eta_p$ to denote the standard Feynman-Kac time marginal measures associated with the NS-SMC algorithm and quantities with an $N$ superscript to correspond to the (adaptive) $N$-particle mean field interpretation of this sequence consistent with the multilevel splitting algorithm described within \citet{Cerou2016} which is algorithmically equivalent, up to the particular choice of Markov kernels employed, to a simple adaptive form of the NS-SMC algorithm if the likelihood is used as the reaction coordinate (the primary difference lies within the particular estimators of interest). Connecting this to the algorithm under study involves the identification of $\eta_p$ with the measure whose Lebesgue density is provided by \eqref{eq:constrained}  and $\gamma_p$ with its unnormalised analogue, $$\gamma_p(d\v x) = \mathcal{P}_p \eta_p(d\v x) = \eta(d\v x) \mathbb{I}\{\mathcal{L}(\v x) > l_p \},$$
and noting the notational incongruity that $\eta_p$ in the present paper corresponds to $\eta_{p+1}$ in \citet{Cerou2016} (but this presents no additional difficulties).

Allow
\begin{align*}
  \phi^{a,b}(\cdot) =& \mathcal{L}(\cdot)  \mathbb{I}_{\mathcal{L}^{-1}\left((a,b]\right)}(\cdot), \\
 \text{and} \quad  \mathcal{Z}^{N,{T^N}} =& \sum_{t=0}^{T^N} \gamma_{t}^N\left(\phi^{L^N_{t},L^N_{t+1}}\right) = \sum_{t=0}^{T^N} {\cal Z}_t^N,
\end{align*}
where $L_0^N = 0$, $L_{T^N+1}^N = \infty$ and for $t=1,\ldots,T^N$, $L_t^N$ is the $(1-\alpha)$ quantile (in the sense of \eqref{eq:ordering}) of $\eta_{t}^N \circ \mathcal{L}^{-1}$ and $L_t$ is the corresponding quantile of $\eta_t \circ \mathcal{L}^{-1}$. 

Let the sequence of ratios of integrated likelihood $\xi_t$ and $\xi_t^N$ be defined, for $t \in \mathbb{N}$, as 
\begin{align*}
  \xi_t^N =& \frac{ \gamma_{t-1}^N\left(\phi^{L^N_{t-1},\infty}\right) }{\gamma_{t-1}^N\left(\phi^{L^N_{t-1},\infty}\right)  + \sum_{p=0}^{t-1} \gamma_p^N\left(\phi^{L^N_{p-1},L^N_p}\right) }\\
  \xi_t =& \frac{\gamma_{t-1}\left(\phi^{L_{t-1},\infty}\right) }{\gamma_{t-1}\left(\phi^{L_{t-1},\infty}\right) + \sum_{p=1}^{t-1} \gamma_p\left(\phi^{L_{p-1},L_p}\right) } = \frac{\gamma_{t-1}\left(\phi^{L_{t-1},\infty}\right)}{\sum_{p=0}^T \mathcal{Z}_p}
\end{align*}
and define 
$T^N = \inf\{t \in \mathbb{N}:\xi_{t}^N < \epsilon \}$  and $T =  \inf\{t \in \mathbb{N}:  \xi_{t} <  \epsilon \}$, 

The proof here is based upon the argument of \cite{Cerou2016} which employs a different stopping criterion to that used here. In order to minimise repetition and duplication of existing arguments we set the threshold $L^*$ in the notation of \cite{Cerou2016} to the $\eta$-essential supremum of the likelihood function so that the algorithm described in that paper never halts. The argument which follows makes use of only intermediate quantities from the proof and deals with the termination of the NS-SMC algorithm explicitly.

\begin{proof}
  We begin by noting that the assumed $\eta$-integrability of the likelihood function is sufficient to ensure that $T$ is finite via Lemma~\ref{lem:terminate}.  Next, by direct application of \cite[Theorem 3.1]{Cerou2016} we have:
  \begin{align}
    L_p^N \cas & L_p \quad \textrm{ for } p=1,\ldots,T \\
    \eta_p^N (\varphi) \inprob & \eta_p(\varphi) \qquad \forall \varphi \in L^2(\mu).
  \end{align}
 By writing $\gamma_p^N(\varphi) = \eta_p^N(\varphi) \prod_{k=1}^{p} \eta_k^N(\mathbb{I}_{\mathcal{L}^{-1}([L_k^N,\infty))})$ we can invoke Lemma~\ref{lemma:adaptive} together with a continuous mapping argument to yield, for all $\varphi \in L^2_c(\mu)$:
  \begin{equation}
    \gamma_p^N(\varphi) \inprob \gamma_p(\varphi). \label{eq:asconv_gammap}
  \end{equation}
  As we have $L_p^N \cas L_p$ for each $p$, we have, noting in the first case that the constant unit function is in ${L^2_c}(\eta_p)$,  by Lemma~\ref{lemma:adaptive} that
  \begin{align}
    \eta_p^N\left(\mathbb{I}_{\mathcal{L}^{-1}((L_p^N,\infty))} \right) \inprob \, & \eta_p\left(\mathbb{I}_{\mathcal{L}^{-1}((L_p,\infty))} \right), \notag\\
    \textrm{ and }\qquad \eta_p^N\left(\phi^{L^N_{p-1},L^N_p}\right) \inprob \, & \eta_p(\phi^{L_{p-1},L_p}). \label{eq:etaconv}
  \end{align}
Writing
  \begin{equation*}
    \gamma_p^N(\varphi) = \eta_p^N(\varphi) \prod_{k=1}^{p} \eta_k^N(\mathbb{I}_{\mathcal{L}^{-1}((L_k^N,\infty))}),
    \end{equation*}
  it follows by a continuous mapping argument that
  \begin{align}
    \gamma_p^N( \phi^{L^N_{p-1},L^N_p} ) \inprob & \gamma_p(\phi^{L_{p-1},L_p}) \label{eq:zconv},
  \end{align}
  and hence (\ref{eq:Zconv}) follows.   Noting that $\{\xi_t\}_t$ is increasing in $t$ and by hypothesis $\xi_{T-1}\neq1-\epsilon$ and $\xi_T = \inf \{ t: \xi_t > 1-\epsilon\}$, there is no $t$ for which $\xi_t=1-\epsilon$ and Lemma~\ref{lemma:randomtime} suffices to ensure that $T^N \cas T$ and hence (\ref{eq:randomtimeconv}) holds. Lemma~\ref{lemma:stopping} then yields (\ref{eq:ZTconv}).  In order to establish the final claim, it is convenient to start from the representation given in \eqref{eq:NSSMCweights}:
  \begin{align*}
    \pi^N(\varphi) = \frac{1}{\sum_{s=1}^{T^N} {\cal Z}_s^N} \sum_{t=0}^{T^N} {\cal Z}_t^N \frac{\eta_t^N(\varphi \cdot \phi^{L_{p-1}^N,L_p^N})}{\eta_t^N(\phi^{L_{p-1}^N,L_p^N})}.
  \end{align*}
  By (\ref{eq:etaconv}) the denominator of the innermost fraction converges in probability, for each $t$, to $\eta_t(\phi_{L_{p-1},L_p})$ and the same argument (up to the replacement of ${\cal L}(\cdot)$ with $\varphi(\cdot) {\cal L}(\cdot)$)  ensures the corresponding convergence of the numerator. Convergence in probability of the individual ${\cal Z}_t^N$ terms to ${\cal Z}_t$  is provided by (\ref{eq:zconv}). Combining these results with Lemma~\ref{lemma:stopping} and the continuous mapping theorem we obtain  \eqref{eq:piconv}.
  \end{proof}
   
\begin{lemma}\label{lem:terminate}
${\cal L} \in L^1(\eta) \Rightarrow T<\infty.$\end{lemma}
\begin{proof} Write $\mathcal{Z}_{t-1} := \int  {\cal L}\, \mathbb{I}_{\{L_{t-1} < \mathcal{L} \le L_t\}}{\rm d}\eta$, and $\mathcal{R}_t := \int  {\cal L}\, \mathbb{I}_{\{\mathcal{L} > L_{t}\}}{\rm d}\eta$. Note that, for $t>1$,
$\mathcal{R}_t = \sum_{k=t+1}^\infty \mathcal{Z}_k < \mathcal{Z} :=  \sum_{k=0}^\infty \mathcal{Z}_k = \int \cal L \, {\rm d}\eta < \infty$, where the first and second inequalities hold by positivity of ${\cal L}$ and hypothesis, respectively. 
Combining that $\mathcal{Z}_t \le \mathcal{Z} < \infty$ with the implication $\sum_{t \in \mathbb{N}} \mathcal{Z}_t < \infty \Rightarrow \mathcal{Z}_t \rightarrow 0$ (e.g., \citet[Thm. 3.22]{rudin1964principles}), we have that $\mathcal{Z}_t \rightarrow 0$.
 Next, observe that
 $\lim_{t\rightarrow \infty}{\cal R}_{t} = {\cal Z} - \lim_{t\rightarrow \infty}\sum_{k=0}^{t}{\cal Z}_k = 0$, 
 where ${\cal R}_t\searrow 0$ (again, due to positivity of ${\cal L}$).
For fixed $\epsilon >0$,
$$
 T<\infty \ \ \Longleftrightarrow \ \ \exists \, t\in \mathbb{N} \quad {\rm s.t.}\quad \frac{\mathcal{R}_{t}}{\mathcal{R}_{t} + \sum_{p=0}^{t-1}\mathcal{\mathcal{Z}}_t} < \epsilon \ \  \Longleftrightarrow  \ \  \exists\, t\in \mathbb{N} \quad {\rm s.t.}\quad \frac{{\mathcal{R}}_t}{\mathcal{Z}} < \epsilon.
$$
As $\mathcal{R}_t \searrow 0$ and $\mathcal{Z}>0$, a suitable $t$ can always be chosen to make $\mathcal{R}_t/\mathcal{Z}$ arbitrarily small, and the result follows.
\\ 
\end{proof}

\subsection{Technical Lemmata}
Throughout this section, all stochastic quantities are assumed to be defined on some common probability space, $(\Omega,\mathcal{F},\mathbb{P})$. In the application of these results, this will be the space upon which a sequence of interacting particle systems of increasing size is defined. We will allow $\mathcal{B}(E)$ to denote the collection of bounded measurable functions on some standard Borel space $(E,\mathcal{E})$.
\begin{lemma}\label{lemma:ctsconv}
Let $\varphi:E \to \mathbb{R}$ and $S:E\to\mathbb{R}$ be continuous. Define the function $$\varphi^a(x) =  \varphi(x) \mathbb{I}_{(a,\infty)}(S(x)),$$ for any $a \in [0,\infty)$. If the sequence of random variables $\{A_N\}_{N\in\mathbb{N}}$ converge almost surely to some constant $a$ then, with probability one, $\varphi^{A_N}$ converges continuously to $\varphi^a$ on $E\setminus S^{-1}(a)$.
\end{lemma}
\begin{proof}
Fix $x \not\in S^{-1}(a)$. Take $a^\prime = (a + S(x))/2$. As $S(x)\neq a$ we have two cases to consider; throughout we restrict ourselves to the event of full probability on which $A_N \to a$.

\emph{Case 1:} $S(x) < a^\prime < a $:
By the convergence of $A_N$ to $a$, there exists $N_0$ such that for all $N > N_0, A_N > a'$. Hence for all $N > N_0$ we have that $\varphi^{A_N}(x^\prime) = 0$ for all $x^\prime \in S^{-1}((0,a^\prime))$. Consequently, $\varphi^{A_N}(x_n) \to \varphi^a(x)$ for any $x_n \to x$ (as the tail of any such sequence is eventually contained within the neighbourhood $S^{-1}((0,a^\prime))$). 

\emph{Case 2:} $S(x) > a^\prime > a $:
By the convergence of $A_N$ to $a$, there exists $N_0$ such that for all $N > N_0, A_N < a'$. Hence for all $N > N_0$ we have that $\varphi^{A_N}(x^\prime) = \varphi^a(x^\prime)$ for all $x^\prime \in S^{-1}((a^\prime,\infty))$. This is a neighbourhood of $x$ and $\varphi^a(x^\prime)$ is itself continuous on this set, continuous convergence follows directly. The result follows as $x \in E \setminus S^{-1}(a)$ was arbitrary.
\end{proof}

\nnew{The next result will be used to show that the convergence of the empirical quantiles of the likelihood function can be transferred to the appropriate quantities within the NS-SMC estimator. We have chosen to use general arguments which show that the empirical measures involved convergence in a weak sense and the functions involved in the estimator are sufficiently regular that standard arguments can be used to verify the convergence. As noted by an anonymous referee, one could establish essentially the same result using elementary approximation arguments along the lines of \citet[Propositions~6.1--6.2]{Cerou2016}.}

\begin{lemma}\label{lemma:adaptive}
Let  $\mu$ be a probability measure on $(E,\mathcal{E})$.  Take a sequence of random probability measures, $\{\mu^N\}_{N \in \mathbb{N}}$ on $(E,\mathcal{E})$ such that, for all $\varphi \in \mathcal{B}(E)$, we have, as $N\to\infty$
  \begin{equation}
    \mu^N(\varphi) \overset{\textrm{a.s.}}{\to} \mu(\varphi) \label{eq:aspointwise},
  \end{equation}
  and for all 
  $\varphi \in L^2(\mu)$ we have, as $N \rightarrow \infty$, 
  \begin{equation}
    \mu^N(\varphi) \inprob \mu(\varphi).
  \end{equation}

Allow $\{A^N\}_{N\in \mathbb{N}}$ and $\{B^N\}_{N\in \mathbb{N}}$ to denote two sequences of random variables that converge almost surely to constants $a$ and $b$, respectively.
  Let $\phi^a = \varphi \cdot \mathbb{I}_{(a,\infty]}(S(\cdot))$
  and $\phi^{a,b} = \varphi \cdot \mathbb{I}_{(a,b]}(S(\cdot))$ for some $\varphi \in L^2_c(\mu)$,
   where $S:E \to [0,\infty)$ is continuous (and hence appropriately measurable) and is such that:
  \begin{equation}\label{eq:likelihood}
    \mu(S^{-1}(\{a,b\})) = 0.
  \end{equation}

Then,
    \begin{align}
       \mu^N(\phi^{A^N}) & \inprob \mu(\phi^a)\label{eq:pconv1} \\
        \mu^N(\phi^{A^N,B^N}) &\inprob \mu(\phi^{a,b}). \label{eq:pconv2} 
    \end{align}
  \end{lemma}

  \begin{proof}
    Writing 
     \begin{equation}\label{eq:vdcdecomp}
         \mu^N(\phi^{A^N}) = \mu^N(\phi^{A^N} \cdot \mathbb{I}_{S^{-1}([0,a))}) + \mu^N(\phi^{A^N} \cdot \mathbb{I}_{S^{-1}(a)}) + \mu^N(\phi^{A^N} \cdot \mathbb{I}_{S^{-1}((a,\infty])}),
     \end{equation}
     it is sufficient to establish that the central term is asymptotically negligible (as $S^{-1}(a)$ is $\mu$-null) whereas the other two converge as required.

     We first show that the central term is asymptotically negligible:
     Note $\mu^N(\phi^{A^N} \cdot \mathbb{I}_{S^{-1}}(a)) \leq \mu^N(|\varphi| \cdot \mathbb{I}_{S^{-1}}(a))$, and that this upper bound converges to zero in probability by hypothesis (as $S^{-1}(a)$ is $\mu$-null and $|\varphi| \in {{L^2}}(\mu)$). In order to establish the convergence of the remaining terms in \eqref{eq:vdcdecomp}, we begin by noting that \eqref{eq:aspointwise} is sufficient to ensure the almost sure weak convergence of $\mu^N$ to $\mu$ via a standard countable determining class argument (see, for example \citep[Theorem 4]{schmon2018large} and the associated remarks). Furthermore, from the definition of $\phi^{A^N}$, we have by Lemma~\ref{lemma:ctsconv} that, with probability one $\phi^{A_N}(x) \to \phi^{a}(x)$ \emph{continuously} on the complement of $S^{-1}(a)$.
    
 As $\varphi \in L^2(\mu)$,  and $\mu$ is a probability measure, \nnew{it is immediate from Jensen's inequality that} $\mu(|\varphi|) < \infty$. By definition, $\phi^{A_N}(\mathbf{x}) \leq |\varphi(\mathbf{x})|$  for all $\mathbf{x}$. By hypothesis, $\mu^N(|\varphi|) \to \mu(|\varphi|) < \infty$ in probability. Hence, for any subsequence $N_m$ there exists a further subsequence $N_{m(k)}$ for which that convergence holds with probability one (see, for example, \citet[Thm 9.2.1]{dudley74real}) .
     As the preimage under a continuous function of an open sets is open and hence a $G_\delta$ set (i.e., expressible as the intersection of countably many open sets), both $S^{-1}([0,a))$ and $S^{-1}((a,\infty))$ viewed as subsets of $E$ with the appropriate subset topologies are themselves Polish spaces by Alexandroff's Theorem \citep{Alexandroff1924}. The almost sure weak convergence of $\mu^N$ to $\mu$ is sufficient to ensure that the restriction of $\mu^N$ to each of these spaces converges weakly to the corresponding restriction of $\mu$ (almost surely). Thus, the conditions of the dominated convergence theorem for vaguely converging measures holds on a set of probability one and we can invoke \cite[Theorem 3.3]{serfozo1982convergence} to establish almost sure convergence of the first and last terms in \eqref{eq:vdcdecomp} to the desired limits.  As for any subsequence $\mu^{N_m}(\phi^{A^{N_m}} \cdot \mathbb{I}_{E \setminus S^{-1}(a)})$ there exists a further subsequence $\mu^{N_{m(k)}}(\phi^{A^{N_{m(k)}}} \cdot \mathbb{I}_{E \setminus S^{-1}(a)})$ which converges almost surely to $\mu(\phi^a \cdot \mathbb{I}_{E \setminus S^{-1}(a)})$ we have that $\mu^{N}(\phi^{A^N} \cdot \mathbb{I}_{E \setminus S^{-1}(a)})$ converges to $\mu(\phi^a \cdot \mathbb{I}_{E \setminus S^{-1}(a)})$ in probability (again, see, for example, \citet[Thm 9.2.1]{dudley74real}) and the first claim follows.  The second claim may be established by writing $\phi^{a,b} = \phi^a - \phi^b$ and applying the first result twice.
 \end{proof}

   \begin{lemma}\label{lemma:randomtime}
     Given the random variables $\{\xi^N_t\}_{N \in \mathbb{N}, t\in \mathbb{N}}$ where, for each t, $\xi^N_t\inprob \xi_t$ for some sequence $\xi_t$, define $T^N = \inf\{t :  \xi_{t}^N > 1 - \epsilon \}$  and $T =  \inf\{t : \xi_{t} > 1 - \epsilon \}$. Provided that $T < \infty$ and $\xi_t < 1 - \epsilon$ for all $t < T$ (i.e. $\xi_t \neq 1 - \epsilon$ for any $t \leq T$) we have that $$T^N \cas T.$$
         \end{lemma}
         \begin{proof}
           Let $A_t^N = \{ \omega : T^N(\omega)  \leq  t \}$ for any $t \in \mathbb{N}$. For any $t < T$, as $ \xi_t < 1- \epsilon$ there exists $\delta_t > \epsilon$ such that $\xi_t < 1 - {\delta_t}$  and as $\xi^N_t \inprob \xi_t$ we have $\lim_{N \to \infty} \mathbb{P}(\xi^N_t > 1- \delta_t) = 0$. Hence $\lim_{N \to \infty} \mathbb{P}(A_t^N) = 0$ for all $t <T$ and $\lim_{N \to\infty} \mathbb{P}(\cup_{t < T} A_t^N) = 0$. 

           Similarly there exists $\delta^\prime_T$ such that $\xi_T > 1 - \delta^\prime_T > 1 - \epsilon$ and 
 $\lim_{N \to \infty} \mathbb{P}(\xi^N_t < 1- \delta_t) = 0$, so $\lim_{N \to \infty} \mathbb{P}(A_T^N) = 1$.             As $\{\omega:T^N(\omega) = T\} = A_T^N \setminus \bigcup_{t<T} A_t^N$, the result follows.
         \end{proof}

 \begin{lemma}\label{lemma:stopping}

   Given some Polish space equipped with its Borel $\sigma$-algebra $(E,\mathcal{E})$, allow $\mathcal{M}(E)$ to denote the collection of measures over $(E,\mathcal{E})$ and equip it with the $\sigma$-algebra generated by bounded measurable functions. For each $t \in \mathbb{N}$, allow $\eta_t^N$ to denote a collection of random measures in $\mathcal{M}(E)$ indexed by $N$, and $\eta_t$ some element of $\mathcal{M}(E)$. For each $t$ in $\mathbb{N}$, let $\varphi_t^N$ denote a collection of random measurable functions from $(E,\mathcal{E})$ to $\mathbb{R}$ indexed by $N$ and  allow $\varphi_t$ to denote some such measurable function.
   
   If for every $t \in \mathbb{N}$, we have $\eta_t^N(\varphi_t^N) \inprob \eta_t(\varphi_{t})$ and $\{T^N\}_{N \in \mathbb{N}}$ is a sequence of $\mathbb{N}$-valued random elements such that $T^N\cas T$, then
   \begin{equation}
     \sum_{s=1}^{T^N} \eta_s^N(\varphi_s^N)  \inprob  \sum_{s=1}^{T} \eta_s(\varphi_{s}).
   \end{equation}

 \end{lemma}
 \begin{proof}
   By the hypothesis of the lemma and the continuous mapping theorem:
 \begin{equation}\label{eq:detsumconv}
     \sum_{s=1}^{T} \eta_s^N(\varphi_s^N)  \inprob  \sum_{s=1}^{T} \eta_s(\varphi_s).
   \end{equation}
   As $T^N \cas T$, there exists, with probability 1, an $N^\star$ such that, for all $N > N^\star$, $T^N=T$ and
     \begin{equation}\label{eq:randsumconv}
     \sum_{s=1}^{T^N} \eta_s^N(\varphi_s^N)  -  \sum_{s=1}^{T} \eta_s^N(\varphi_s^N) = 0,
   \end{equation}
   telling us that the left hand side of \eqref{eq:randsumconv} converges almost surely to zero and so:
   \begin{equation}\label{eq:randsumconv2}
     \sum_{s=1}^{T^N} \eta_s^N(\varphi_s^N)  -  \sum_{s=1}^{T} \eta_s^N(\varphi_s^N) \inprob 0.
   \end{equation}
   Adding the left hand sides of \eqref{eq:detsumconv} and \eqref{eq:randsumconv2} and, recalling that the limit in probability of the element-wise sum of two sequences which converge in probability is the sum of their respective limits, the claim follows. 
  \end{proof}

\section{\new{Empirical Results with Stratified Resampling}}

\new{This appendix presents results when stratified resampling is used in place of multinomial resampling. This affects SMC methods. The performance of NS and of the spike-and-slab methods using exact sampling are not affected by this change.}

\new{Results for the spike-and-slab example are shown in Table \ref{tbl:sphere10_stratified}. NS-SMC is still the best-performing method in terms of mean squared error. The null hypothesis of unbiasedness is rejected for the same combinations of method and $N$, plus now also for ATA-SMC with $N=10^3$.}

\begin{figure*}[ht!]
  \centering
   \includegraphics[angle=360, width=0.8\textwidth]{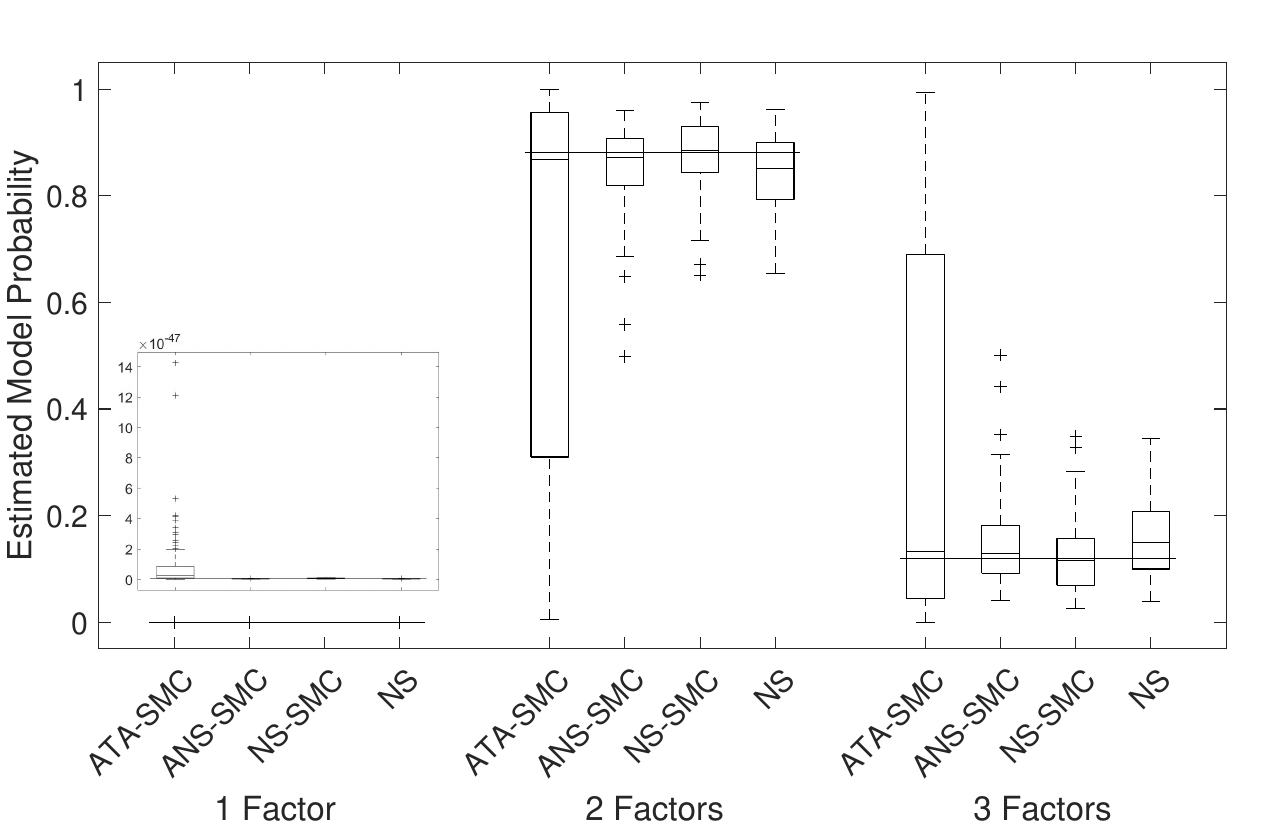}
  \caption{\new{Stratified resampling alternative to Figure 2.}}\label{fig:boxFA_stratified}
\end{figure*}

\begin{figure}[H]
	\centering \small
	\subfloat[ATA-SMC]{\includegraphics[width=0.32\textwidth]{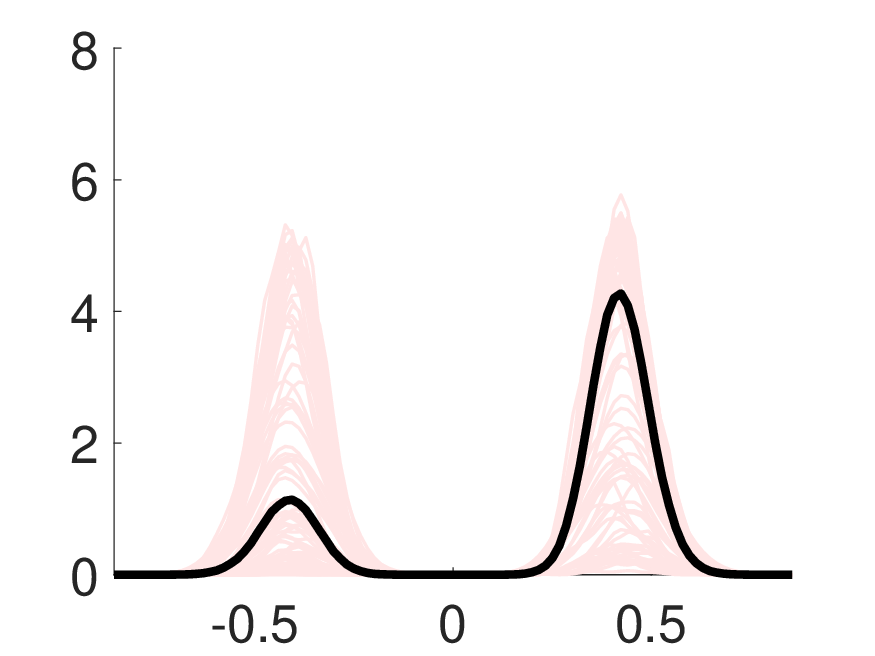}}
	\subfloat[NS]{\includegraphics[width=0.32\textwidth]{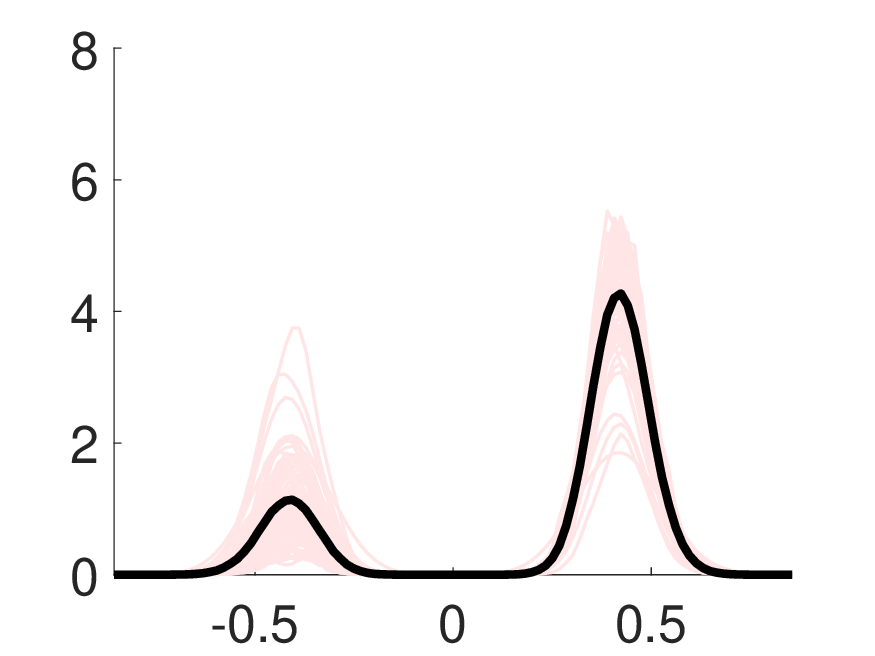}} 
	\subfloat[ANS-SMC]{\includegraphics[width=0.32\textwidth]{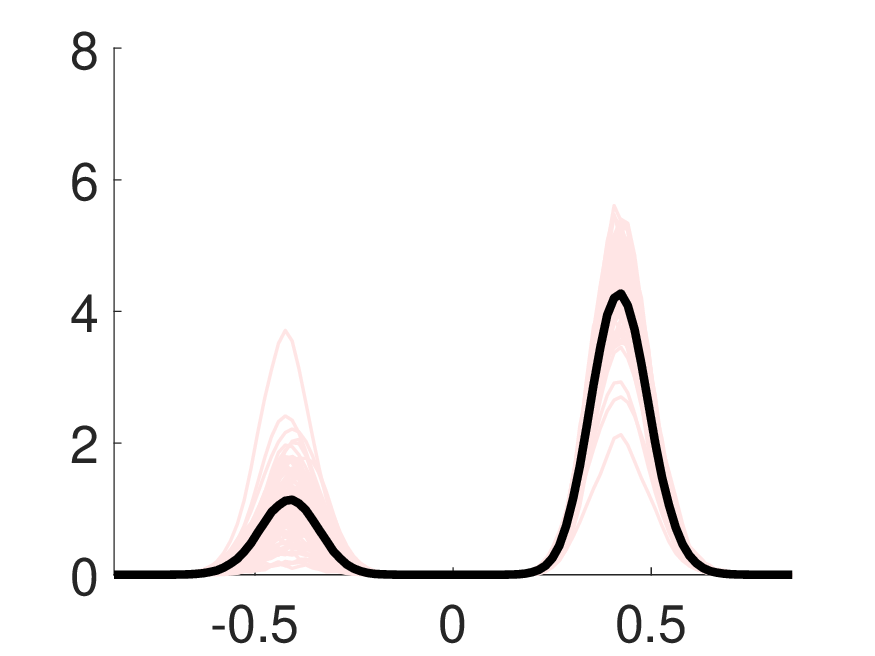}}
	\caption{\new{Stratified resampling alternative to Figure 3.}}
	\label{fig:FA2_marginals_stratified}
\end{figure}

\new{In the factor analysis example, Figures \ref{fig:boxFA_stratified} and \ref{fig:FA2_marginals_stratified} are remarkably similar to their multinomial resampling counterparts. The relative average KSD values for the factor analysis example (Table \ref{tab:FA_KSD_stratified}) have changed the most, though ANS-SMC remains competitive.}

\begin{table*}[ht!] \footnotesize
\centering \scriptsize
\caption{\new{Stratified resampling alternative to Table 2 \nnew{in the paper}.}}
\label{tbl:sphere10_stratified}
\begin{tabular}{c|c|c|c|c|c|c|c}
 \multicolumn{2}{c}{}    & \multicolumn{2}{c}{$N=10^2$}  & \multicolumn{2}{c}{$N=10^3$}    & \multicolumn{2}{c}{$N=10^4$} \\ 
 \multicolumn{2}{c}{}    & \multicolumn{2}{c}{($10^4$ repeats)}  & \multicolumn{2}{c}{($10^3$ repeats)}    & \multicolumn{2}{c}{($10^2$ repeats)} \\ 
sampler   & method   & ${\cal Z}^{N}$(SE) & \multicolumn{1}{c|}{evals} & ${\cal Z}^{N}$(SE) & \multicolumn{1}{c|}{evals} & ${\cal Z}^{N}$(SE)       & evals      \\ \hline 
 Exact & NS & \textcolor{red}{0.4532 (0.0026)} & \num{1.0e+4} & 0.3974 (0.0021) & \num{1.0e+5} & \textbf{0.3913 (0.0019)} & \num{1.0e+6} \\ 
 & $\text{NS}^\star$ & \textbf{0.3866 (0.0023)} & \num{1.0e+4} & \textbf{0.3912 (0.0021)} & \num{1.0e+5} & 0.3907 (0.0019) & \num{1.0e+6} \\  
 & ANS-SMC & 0.3953 (0.0033) & \num{5.1e+3} & 0.3942 (0.0028) & \num{5.0e+4} & 0.3931 (0.0027) & \num{5.0e+5} \\ 
 & NS-SMC & 0.3927 (0.0031) & \num{1.0e+4} & 0.3940 (0.0028) & \num{1.0e+5} & 0.3969 (0.0031) & \num{1.0e+6} \\ 
MCMC & NS & \textcolor{red}{0.6235 (0.0234)} & \num{1.0e+5} & \textcolor{red}{0.4136 (0.0067)} & \num{9.9e+5} & 0.4034 (0.0066) & \num{9.8e+6} \\  
 & $\text{NS}^\star$  & \textcolor{red}{0.5346 (0.0203)} & \num{1.0e+5} & 0.4071 (0.0066) & \num{9.9e+5} & 0.4028 (0.0066) & \num{9.8e+6} \\ 
 & ANS-SMC  & \textcolor{red}{0.4500 (0.0072)} & \num{5.0e+4} & 0.4006 (0.0044) & \num{4.9e+5} & 0.3907 (0.0044) & \num{4.9e+6} \\ 
 & NS-SMC  & \textbf{0.3954 (0.0053)} & \num{1.0e+5} & \textbf{0.3908 (0.0041)} & \num{9.9e+5} & \textbf{0.3936 (0.0040)} & \num{9.8e+6} \\ 
 & ATA-SMC & 0.2687 (0.1523) & \num{1.7e+5} & \textcolor{red}{0.0844 (0.0155)} & \num{1.7e+6} & \textcolor{red}{0.0691 (0.0135)} & \num{1.8e+7} \\ 
 & TA-SMC & \textcolor{red}{0.1808 (0.0524)} & \num{3.4e+5} & \textcolor{red}{0.0691 (0.0139)} & \num{3.5e+6} & 0.4773 (0.2137) & \num{3.6e+7} \\ 
\end{tabular}
\end{table*}

\begin{table}[ht!]
  \centering 
  \caption{\new{Stratified resampling alternative to Table 3. }} \small
  \label{tab:FA_KSD_stratified}
  \begin{tabular}{c|c|c|c|c}
 Factors & Method & Avg. KSD & Avg. evals & Avg. time\\ \hline 
1 & ATA-SMC  & 1.54  & \num{4.0e5} & \num{4.6e1} \\ 
    & ANS-SMC   & 1.24  & \num{3.5e5} & \num{5.0e1} \\ 
    & NS-SMC   & 1.30 & \num{6.9e5} & \num{9.8e1} \\ 
    & NS    &  1.36   & \num{4.1e5} & \num{9.8e1} \\ 
\hline 
2 & ATA-SMC   & 5.36 & \num{8.9e5} & \num{1.4e2} \\ 
    & ANS-SMC  & 3.58  & \num{6.8e5} & \num{1.5e2} \\ 
    & NS-SMC  & 3.71 & \num{1.4e6} & \num{2.9e2} \\ 
    & NS  & 4.14 & \num{7.9e5} & \num{2.9e2} \\ 
\hline
3 & ATA-SMC & 7.82   & \num{1.3e6} & \num{3.6e2} \\ 
    & ANS-SMC  &  6.90  & \num{1.0e6} & \num{4.4e2} \\ 
    & NS-SMC   & 9.29  & \num{2.1e6} & \num{8.8e2}\\ 
    & NS         &  7.54  & \num{1.2e6} & \num{7.1e2}\\ 
\end{tabular}
\end{table}

\section{\nnew{Comparison of Weight Choices in Nested Sampling}}\label{ap:CompareWeight}

Here we examine further, in the original (idealised) NS setting, the choice of $\exp(-t/N)$ versus that of $((N-1)/N)^t$ as increasing amounts of prior exploration are needed to access the bulk of the integral. Firstly, we study the behaviour of the accuracy of estimates of $p_t$ on a problem where analytical quantiles are available. Specifically, we set $\eta(x) = \mathbb{I}\{x \in (0,1) \}$, i.e., uniform on the interval $(0,1)$, and set 
$$\mathcal{L}(x)=0.1(1-x) +  1.9\cdot \mathbb{I}\{x<v\}\frac{(v-x)}{v^2},$$
for $v \in (0,1)$. By construction, both $(1-x)$ and  $\mathbb{I}\{x<v\}\frac{(v-x)}{v^2}$ integrate to $1/2$ under $\eta$ (thus, $\mathcal{Z} = 1$ for any choice of $v$), and ninety-five percent of the integral lies in $(0, v)$, again for any choice of $v$). 

As $\mathcal{L}$ is strictly monotonically decreasing in $x$, the associated constrained distribution is 
$$\eta(x ; \mathcal{L}(\breve{x})) \propto \mathbb{I}\{x \in (0, \breve{x})\},$$
i.e., uniform on the interval $(0,\breve{x})$, so generating exact samples from the constrained prior is straightforward.  Moreover, the problem setup allows us to track the true value of $p_t$ for an observed value of $x$, as the two values are equal by construction.

\begin{figure}[ht!]
    \centering \includegraphics[width=1\linewidth]{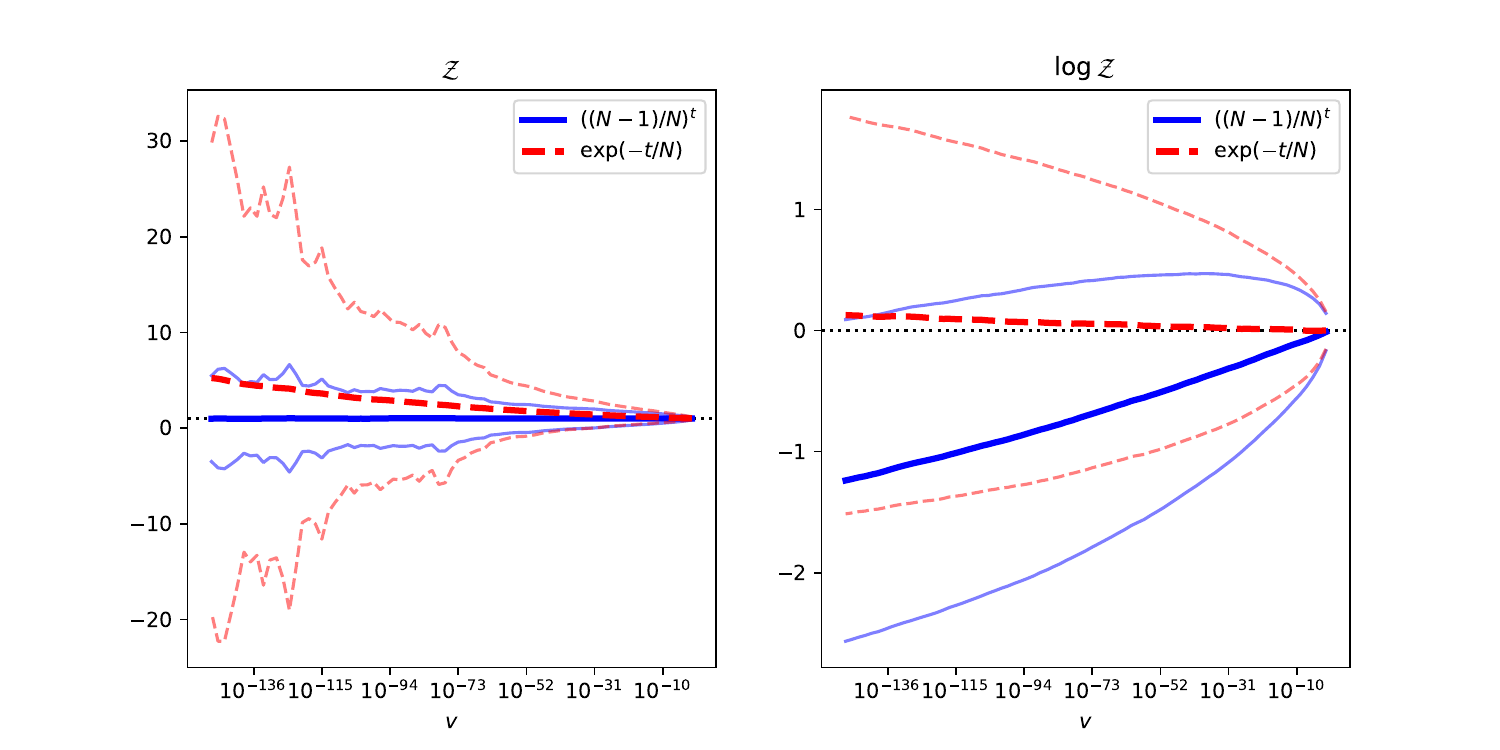}
    \caption{Comparison of weight choices for NS. Thicker lines indicate mean result over the runs, and the thinner lines representing the mean result $\pm$ one standard deviation of the results across the runs.  }
    \label{fig:Zvar}
\end{figure}

The estimator $\exp(-t/N)$ tracks the typical behaviour of the true $p_t$ well.
Obtaining the mean and median $p_t$ values from 1000 simulations (each observed at the same values of $t$) with $N=100$ yields Table \ref{tab:mixture}.  In particular, the choice of $\exp(-t/N)$ tracks the median very well. The value $((N-1)/N)^t$ appears to track {\em neither} the median or even the mean closely, with the accuracy difference compared to $\exp(-t/N)$ becoming larger as iterations increase. However, such a property does not necessarily yield a lower variance estimator when using the weights $\exp(-t/N)$ over $((N-1)/N)^t$. Figure \ref{fig:Zvar} below 
plots the mean value and variability of the observed results.

\begin{table}[ht!] \footnotesize
\centering
\caption{Comparison of estimates of $p_t$ and typical values for increasing $t$ in the mixture example.}
\label{tab:mixture}
\begin{tabular}{c|c|c|c|c|c}
$t$  & $\exp(-t/N)$ & $((N-1)/N)^t$  &  ${\rm mean}(p_t)$   &  ${\rm median}(p_t)$ \\ \hline 
1000 & $4.5 \times 10^{-5}$ & $4.3 \times 10^{-5}$ & $4.8 \times 10^{-5}$ & $4.5 \times 10^{-5}$ \\
3000 & $9.4 \times 10^{-14}$ & $8.1 \times 10^{-14}$ & $1.1 \times 10^{-13}$ & $9.2 \times 10^{-14}$ \\
5000 & $1.9 \times 10^{-22}$ & $1.5 \times 10^{-22}$ & $2.4 \times 10^{-22}$ & $1.9 \times 10^{-22}$ \\
7000 & $4.0 \times 10^{-31}$ & $2.8 \times 10^{-31}$ & $5.6 \times 10^{-31}$ & $3.9 \times 10^{-31}$ \\
9000 & $8.2 \times 10^{-40}$ & $5.2 \times 10^{-40}$ & $1.3 \times 10^{-39}$ & $8.1 \times 10^{-40}$ \\
11000 & $1.7 \times 10^{-48}$ & $9.7 \times 10^{-49}$ & $2.9 \times 10^{-48}$ & $1.7 \times 10^{-48}$ \\
13000 & $3.5 \times 10^{-57}$ & $1.8 \times 10^{-57}$ & $6.6 \times 10^{-57}$ & $3.5 \times 10^{-57}$ \\
15000 & $7.2 \times 10^{-66}$ & $3.4 \times 10^{-66}$ & $1.5 \times 10^{-65}$ & $7.0 \times 10^{-66}$ \\
17000 & $1.5 \times 10^{-74}$ & $6.3 \times 10^{-75}$ & $3.4 \times 10^{-74}$ & $1.5 \times 10^{-74}$ \\
19000 & $3.0 \times 10^{-83}$ & $1.2 \times 10^{-83}$ & $7.8 \times 10^{-83}$ & $3.0 \times 10^{-83}$ \\
21000 & $6.3 \times 10^{-92}$ & $2.2 \times 10^{-92}$ & $1.8 \times 10^{-91}$ & $6.1 \times 10^{-92}$ \\
23000 & $1.3 \times 10^{-100}$ & $4.1 \times 10^{-101}$ & $4.0 \times 10^{-100}$ & $1.3 \times 10^{-100}$ \\
25000 & $2.7 \times 10^{-109}$ & $7.6 \times 10^{-110}$ & $9.3 \times 10^{-109}$ & $2.6 \times 10^{-109}$ \\
27000 & $5.5 \times 10^{-118}$ & $1.4 \times 10^{-118}$ & $2.3 \times 10^{-117}$ & $5.4 \times 10^{-118}$ \\
29000 & $1.1 \times 10^{-126}$ & $2.6 \times 10^{-127}$ & $5.3 \times 10^{-126}$ & $1.1 \times 10^{-126}$ \\
31000 & $2.3 \times 10^{-135}$ & $4.9 \times 10^{-136}$ & $1.2 \times 10^{-134}$ & $2.3 \times 10^{-135}$ \\
33000 & $4.8 \times 10^{-144}$ & $9.2 \times 10^{-145}$ & $2.9 \times 10^{-143}$ & $4.7 \times 10^{-144}$ 
\end{tabular}
\end{table}

\begin{figure}[ht!]
    \centering \includegraphics[width=1\linewidth]{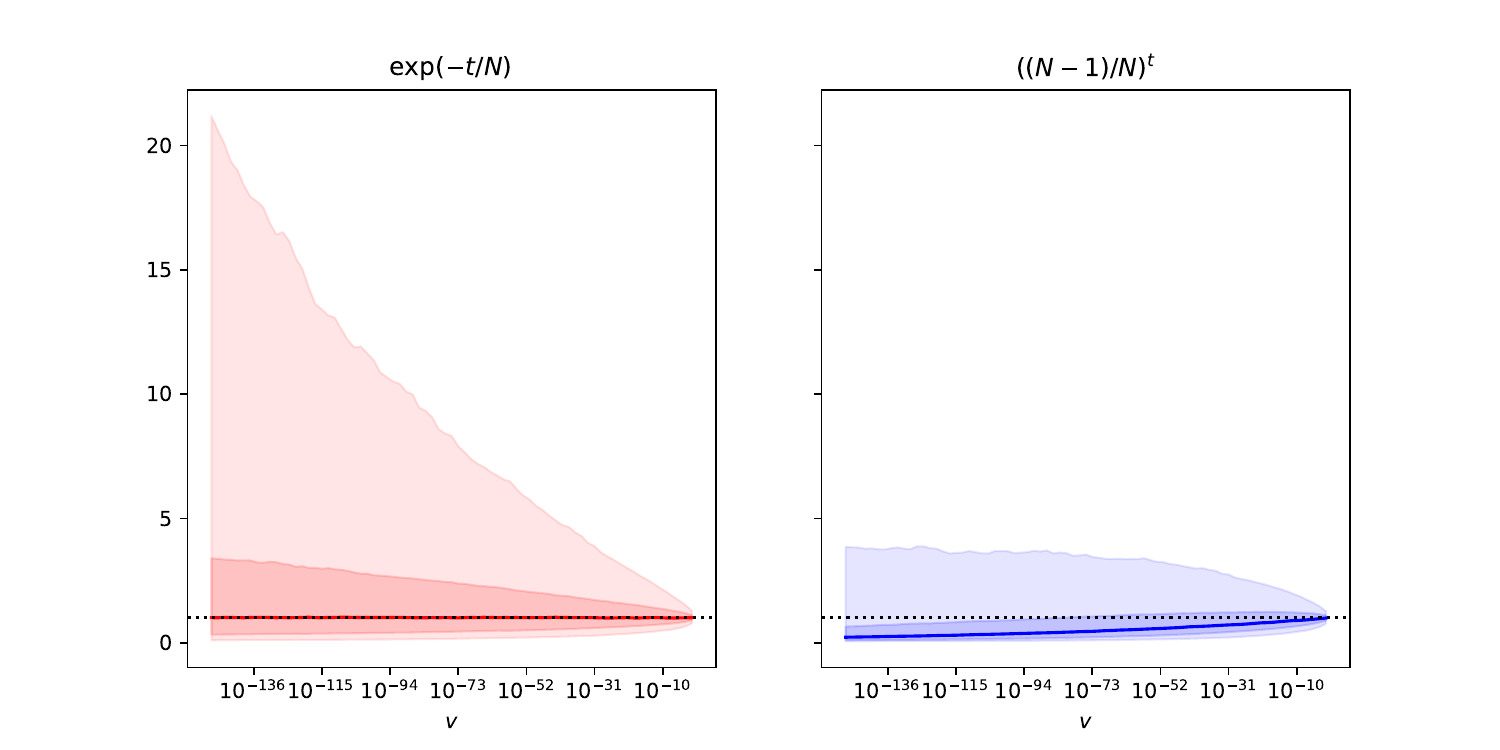}
    \caption{Median and the 0.05, 0.25, 0.5, and 0.95 quantiles of the estimators arising from the two weight choices.}
    \label{fig:Zquantiles}
\end{figure}

The weights $\exp(-t/N)$ tend to yield estimators with median closer to $\mathcal{Z}$ value for this example, as demonstrated in Figure \ref{fig:Zquantiles}. This is not surprising since positive-valued unbiased estimators with sufficient variance must be positively skewed, so one would expect the median of such an estimator to underestimate the true value.

Thus, whilst an estimator constructed with weights $((N-1)/N)^t$ may exhibit lower variance (and bias for $\mathcal{Z}$), it is worth noting that using the one constructed from $\exp(-t/N)$ may potentially be a better pragmatic choice in some settings (assuming that the sampling procedure is sufficiently close to that of idealized setting). Note also that $\lim_{N \rightarrow \infty}\left(\frac{N-1}{N}\right)^t=\exp(-t/N)$, so any differences in estimates will be smaller for larger $N$ and/or smaller $t$.

\end{document}